\documentclass[final, a4paper]{aastex631}

\usepackage{color}
\usepackage{enumitem}
\usepackage{hyperref}
\usepackage{verbatim}
\usepackage{amsmath,amstext,mathrsfs}
\usepackage[all]{hypcap} 
\usepackage{afterpage}    
\usepackage{marginnote}
\usepackage{makecell}
\usepackage{subfigure}
\usepackage{multirow}
\usepackage{amsmath}

\shorttitle{M17 Magnetic Fields}
    \shortauthors{Zhao et al}
\usepackage{enumitem}
\setlist[enumerate]{listparindent=\parindent}
\makeatletter

\newcommand{\Rmnum}[1]{\expandafter\slowromancap\romannumeral #1@}
\makeatother

\begin{document}
    \hfuzz = 150pt
\title{\bfseries BISTRO Survey: Gravity-Dominated and Magnetically Regulated Star Formation in M17 SW}

\author[0000-0003-0596-6608]{Mengke Zhao}
\affiliation{School of Astronomy and Space Science, Nanjing University, 163 Xianlin Avenue, Nanjing 210023, People's Republic of China}
\affiliation{Key Laboratory of Modern Astronomy and Astrophysics (Nanjing University), Ministry of Education, Nanjing 210023, People's Republic of China}

\author[0000-0002-5093-5088]{Keping Qiu}
\affiliation{School of Astronomy and Space Science, Nanjing University, 163 Xianlin Avenue, Nanjing 210023, People's Republic of China}
\affiliation{Key Laboratory of Modern Astronomy and Astrophysics (Nanjing University), Ministry of Education, Nanjing 210023, People's Republic of China}

\author[0000-0001-7379-6263]{Ji-hyun Kang}
\affiliation{Korea Astronomy and Space Science Institute, 776 Daedeokdae-ro, Yuseong-gu, Daejeon 34055, Republic of Korea}

\author[0000-0002-4154-4309]{Xindi Tang}
\affiliation{Xinjiang Astronomical Observatory, Chinese Academy of Sciences, 830011 Urumqi, People's Republic of China}

\author[0000-0002-1178-5486]{Anthony Whitworth}
\affiliation{School of Physics and Astronomy, Cardiff University, The Parade, Cardiff, CF24 3AA, UK}

\author[0000-0003-1140-2761]{Derek Ward-Thompson}
\affiliation{Jeremiah Horrocks Institute, University of Lancashire, Preston PR1 2HE, UK}

\author[0000-0002-8234-6747]{Takashi Onaka}
\affiliation{Department of Astronomy, Graduate School of Science, The University of Tokyo, 7-3-1 Hongo, Bunkyo-ku, Tokyo 113-0033, Japan}

\author[0000-0002-3179-6334]{Chang Won Lee}
\affiliation{Korea Astronomy and Space Science Institute, 776 Daedeokdae-ro, Yuseong-gu, Daejeon 34055, Republic of Korea}
\affiliation{University of Science and Technology, Korea (UST), 217 Gajeong-ro, Yuseong-gu, Daejeon 34113, Republic of Korea}

\author[0000-0001-7491-0048]{Tyler L. Bourke}
\affiliation{SKA Observatory, Jodrell Bank, Lower Withington, Macclesfield SK11 9FT, UK}
\affiliation{Jodrell Bank Centre for Astrophysics, School of Physics and Astronomy, University of Manchester, Oxford Road, Manchester, M13 9PL, UK}

\author[0000-0001-7866-2686]{Jihye Hwang}
\affiliation{Institute for Advanced Study, Kyushu University, Japan}
\affiliation{Department of Earth and Planetary Sciences, Faculty of Science, Kyushu University, Nishi-ku, Fukuoka 819-0395, Japan}

\author{David Eden}
\affiliation{Armagh Observatory and Planetarium, College Hill, Armagh BT61 9DG, UK}

\author[0000-0003-2017-0982]{Thiem Hoang}
\affiliation{Korea Astronomy and Space Science Institute, 776 Daedeokdae-ro, Yuseong-gu, Daejeon 34055, Republic of Korea}
\affiliation{University of Science and Technology, Korea, 217 Gajeong-ro, Yuseong-gu, Daejeon 34113, Republic of Korea}

\author[0000-0002-6510-0681]{Motohide Tamura}
\affiliation{National Astronomical Observatory of Japan, National Institutes of Natural Sciences, Osawa, Mitaka, Tokyo 181-8588, Japan}
\affiliation{Department of Astronomy, Graduate School of Science, The University of Tokyo, 7-3-1 Hongo, Bunkyo-ku, Tokyo 113-0033, Japan}
\affiliation{Astrobiology Center, National Institutes of Natural Sciences, 2-21-1 Osawa, Mitaka, Tokyo 181-8588, Japan}

\author[0000-0003-2815-7774]{Jungmi Kwon}
\affiliation{Department of Astronomy, Graduate School of Science, The University of Tokyo, 7-3-1 Hongo, Bunkyo-ku, Tokyo 113-0033, Japan}

\author{Felix Priestley}
\affiliation{School of Physics and Astronomy, Cardiff University, The Parade, Cardiff, CF24 3AA, UK}

\author[0000-0003-2412-7092]{Kee-Tae Kim}
\affiliation{Korea Astronomy and Space Science Institute, 776 Daedeokdae-ro, Yuseong-gu, Daejeon 34055, Republic of Korea}
\affiliation{University of Science and Technology, Korea (UST), 217 Gajeong-ro, Yuseong-gu, Daejeon 34113, Republic of Korea}

\author{Doris Arzoumanian}
\affiliation{Division of Science, National Astronomical Observatory of Japan, 2-21-1 Osawa, Mitaka, Tokyo 181-8588, Japan}

\author[0000-0002-9289-2450]{James Di Francesco}
\affiliation{NRC Herzberg Astronomy and Astrophysics, 5071 West Saanich Road, Victoria, BC V9E 2E7, Canada}
\affiliation{Department of Physics and Astronomy, University of Victoria, Victoria, BC V8W 2Y2, Canada}

\author[0000-0003-4761-6139]{Chakali Eswaraiah}
\affiliation{Department of Physical Sciences, Indian Institute of Science Education and Research (IISER) Mohali, Knowledge City, Sector 81, SAS Nagar 140306, Punjab, India}

\author[0000-0002-6773-459X]{Doug Johnstone}
\affiliation{NRC Herzberg Astronomy and Astrophysics, 5071 West Saanich Road, Victoria, BC V9E 2E7, Canada}
\affiliation{Department of Physics and Astronomy, University of Victoria, Victoria, BC V8W 2Y2, Canada}

\author[0000-0002-5913-5554]{Nguyen Bich Ngoc}
\affiliation{Vietnam National Space Center, Vietnam Academy of Science and Technology, 18 Hoang Quoc Viet, Hanoi, Vietnam}

\author{Zhiwei Chen}
\affiliation{Purple Mountain Observatory, Chinese Academy of Sciences, 2 West Beijing Road, 210008 Nanjing, People's Republic of China}

\author{Sarah Sadavoy}
\affiliation{Department for Physics, Engineering Physics and Astrophysics, Queen's University, Kingston, ON, K7L 3N6, Canada}

\author[0000-0002-6386-2906]{Archana Soam}
\affiliation{Indian Institute of Astrophysics, II Block, Koramangala, Bengaluru 560034, India}

\author{Ray S. Furuya}
\affiliation{Institute of Liberal Arts and Sciences Tokushima University, Minami Jousanajima-machi 1-1, Tokushima 770-8502, Japan}

\author[0000-0001-5522-486X]{Shih-Ping Lai}
\affiliation{Institute of Astronomy and Department of Physics, National Tsing Hua University, Hsinchu 30013, Taiwan}
\affiliation{Academia Sinica Institute of Astronomy and Astrophysics, No.1, Sec. 4., Roosevelt Road, Taipei 10617, Taiwan}

\author[0000-0003-4022-4132]{Woojin Kwon}
\affiliation{Department of Earth Science Education, Seoul National University, 1 Gwanak-ro, Gwanak-gu, Seoul 08826, Republic of Korea}
\affiliation{SNU Astronomy Research Center, Seoul National University, 1 Gwanak-ro, Gwanak-gu, Seoul 08826, Republic of Korea}

\author[0000-0002-0794-3859]{Pierre Bastien}
\affiliation{Centre de recherche en astrophysique du Qu\'{e}bec \& d\'{e}partement de physique, Universit\'{e} de Montr\'{e}al,1375, Avenue Thérèse-Lavoie-Roux, Montréal, QC, H2V OB3, Canada}

\author[0000-0002-8557-3582]{Kate Pattle}
\affiliation{Department of Physics and Astronomy, University College London, WC1E 6BT London, UK}

\author[0000-0001-6524-2447]{David Berry}
\affiliation{Department of Physics, University of Bath, Claverton Down, Bath, BA2 7AY, United Kingdom}
\affiliation{East Asian Observatory, 660 N. A'oh\={o}k\={u} Place, University Park, Hilo, HI 96720, USA}

\correspondingauthor{Keping Qiu}
\email{kpqiu@nju.edu.cn}



\begin{abstract}

We present high-resolution magnetic field maps of the M17 SW molecular cloud using JCMT 850 µm dust polarization at scale of 14$''$.
The magnetic field exhibits a distinct arc-like structure that encircles three dense clumps (C1, C2, and C3). 
By combining polarization data with ammonia line observations, the plane-of-sky magnetic field strength, measured using the Skalidis-Tassis method to minimize angle dispersion errors, ranges from 0.1 to 2.4 mG (mean: 0.54 mG). 
Energy budget analysis reveals a hierarchy dominated by gravity ($e_G \approx 10^{-7.8}$ erg cm$^{-3}$), which exceeds both magnetic ($e_B \approx 10^{-8.3}$ erg cm$^{-3}$) and turbulent ($e_k \approx 10^{-8.7}$ erg cm$^{-3}$) energies. 
Since all three energy densities lie within one order of magnitude, gravitational dominance acts primarily as the global driver, while the system remains in a state of near-equipartition.
Structurally, the northeastern boundary shows magnetic field lines perpendicular to the shock front, consistent with compression from the adjacent HII region. 
Within the cloud, magnetic field lines generally align with gravity to assist collapse, but turn perpendicular to gravity within curved accretion bridges. 
This configuration provides support against radial collapse while guiding gas flow. Kinematic evidence suggests that these channels transport material from Clump C3 onto the massive Clump C2. 
Star formation in M17 SW is globally driven by gravity but locally regulated by the magnetic field structure.

\end{abstract}

\section{introduction}

Star formation, a fundamental yet complex process governing interstellar medium and galaxy evolution  \citep{1987ARA&A..25...23S,2007ARA&A..45..565M}, is regulated by magnetic field  \citep{2012ARA&A..50...29C,2021Galax...9...41L,2023ASPC..534..193P}, gravity \citep{2019MNRAS.490.3061V,2022MNRAS.514L..16L,2024NatAs...8..472L} and turbulence  \citep{1981MNRAS.194..809L,2021Galax...9...41L,2023ApJ...946L..46C}.
However, their exact roles in the star-formation process are not entirely understood  \citep{2015Natur.520..518L,2012ARA&A..50...29C,2023ASPC..534..193P}. 
A central debate in this field concerns the relative importance of these forces: whether molecular clouds are supported by magnetic fields in a sub-critical state, or whether they are governed by super-critical gravitational collapse and supersonic turbulence \citep{2012ARA&A..50...29C,2021Galax...9...41L,2023ASPC..534..193P}. 
Energetically, this interplay is often quantified by comparing energy densities, such as gravitational potential energy $e_G$ vs. magnetic energy $e_B$  \citep{2018MNRAS.474.2167L} or dimensionless parameters like the mass-to-flux ratio ($\lambda$, \citealt{2004ApJ...600..279C}) and Alfvénic Mach number ($\mathcal{M}_A$, \citealt{2024ApJ...976..209Z}). 
However, scalar metrics alone obscure the geometric (vector) role of magnetic fields.
The morphology of magnetic field, gravity and turbulence is also critical: magnetic field orientation parallel to gas flows offers little resistance to collapse, whereas fields perpendicular to the flow can provide magnetic support, channeling material into dense cores  \citep{2014ApJ...794L..18Q,2015Natur.520..518L,2024arXiv240809690Z}.
This distinction is particularly vital in complex environments where external feedback may reshape the cloud's geometry, potentially decoupling the local B-field orientation from the global energy budget  \citep{2021ApJ...912....2H,2022ApJ...934...45Z}.

\begin{figure}
    \centering
    \includegraphics[width=0.5\linewidth]{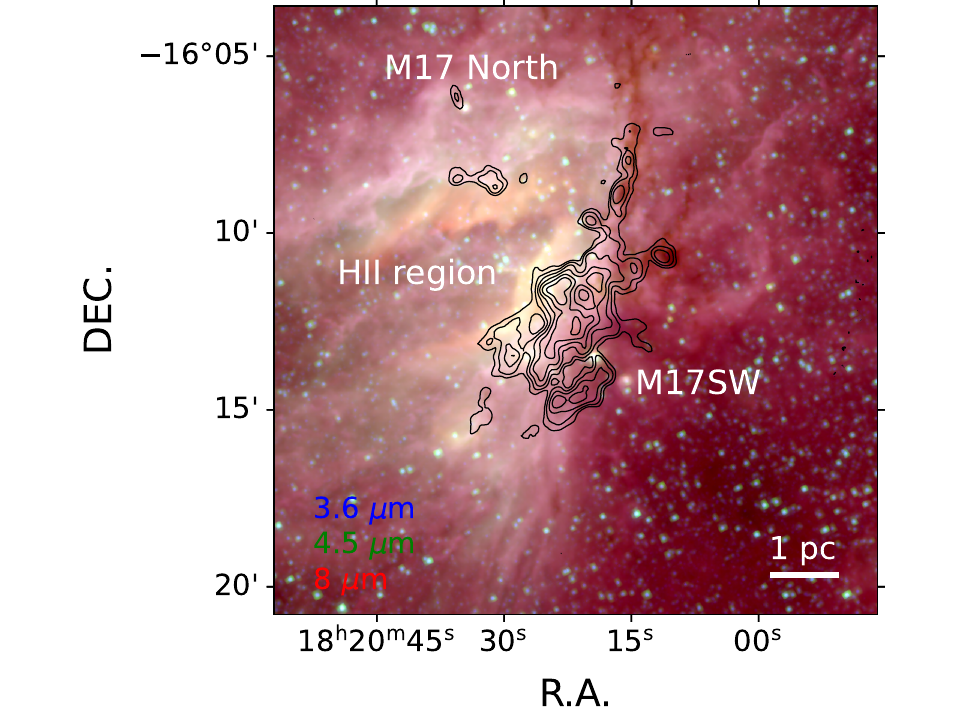}
    \caption{. Large-scale mid-infrared view of the M17 star-forming complex.
    The background image is a three-color composite from Spitzer \citep{2003PASP..115..953B,2009PASP..121...76C} showing the interaction between the HII region and the molecular cloud.
    The bright emission to the bright left/north traces the ionizing cluster NGC 6618 and the photo-dissociation region (PDR).
    The black contours show the coverage of our JCMT BISTRO 850 $\mu$m polarization observations centered on the M17 SW molecular cloud.
Note the sharp boundary facing the HII region, indicative of the strong external compression discussed in the text.}
    \label{figbag}
\end{figure}

The M17 star-forming complex, specifically the M17 SW molecular cloud  \citep{1974ApJ...189L..35L,1997ApJ...489..698H,2009ApJ...696.1278P}, provides an ideal laboratory to test this dual framework.
Located at a distance of 1.98 kpc  \citep{2011ApJ...733...25X,2012PASJ...64..110C}, M17 SW is a prototypical photon-dominated region (PDR) subjected to intense compressive feedback from the adjacent ionizing cluster NGC 6618  \citep{1991ApJ...374..533L,2009ApJ...696.1278P,2020ApJ...888...98L} (see Fig.\,\ref{figbag}). 
The region features extreme column densities and active high-mass star formation, suggesting a system under strong compressive forcing. 
Recent studies have extensively characterized the physical environment of M17, ranging from its dust properties \citep{2025AJ....170..125C} to the dense gas kinematics revealed by ammonia surveys  \citep{2019ApJ...884....4K}.
These works confirm the extreme nature of the region, where accurately quantifying the energy budget (scalar analysis) is essential to determining the stability of the cloud.

The magnetic field morphology of M17 has been the subject of multi-wavelength investigation. 
On large scales, near-infrared polarimetry  \citep{2012PASJ...64..110C} revealed a uniform magnetic field orientation aligned with the Galactic plane, suggesting that the global magnetic field is strong enough to resist turbulent randomization in the diffuse envelope. 
Moving to the far-infrared, observations with SOFIA/HAWC+ have traced the magnetic field into denser regions  \citep{2022ApJ...929...27H}. 
While SOFIA traces the warm dust, resolving the magnetic field geometry in dense cold structures requires the resolution and sensitivity accessible at submillimeter wavelengths.

In this work, we present 850 $\mu$m dust polarization observations of M17 SW using the James Clerk Maxwell Telescope (JCMT), as a part of the B-fields In Star-forming Region Observations (BISTRO, \citealt{2017ApJ...842...66W,2020pase.conf..117B}).
The spatial resolution in this work is significantly higher than Planck  \citep{2020A&A...641A..12P} and similar to that observed by SOFIA at Far-infrared  \citep{2022ApJ...929...27H}.
Due to longer wavelengths, the dust polarization at sub-millimeter wavelength probes the magnetic field at the cold and inner ISM that remains optically thick in the far-infrared. 
Complemented by Green Bank Telescope NH$_3$ $(1,1)/(2,2)$ spectroscopy  \citep{2017ApJ...843...63F,2019ApJ...884....4K} tracing gas kinematics and temperature, we quantify the relative contributions of magnetic fields, gravity, and turbulence through:
(1) A scalar analysis using the physical parameters to establish the global energy hierarchy ($e_G, e_B, e_K$), and (2) a vector analysis of the alignment between magnetic fields, gravitational gradients, and velocity fields. 
This synthesis reveals a physical picture where gravity dictates the global collapse, while the magnetic field, shaped by external compression, regulates the specific route of mass accretion.

\section{Data}

\subsection{Polarization Observations and Reduction}

We present 850 $\mu$m dust polarization observations of M17 SW obtained with the POL-2 polarimeter on the James Clerk Maxwell Telescope (JCMT). 
As part of the B-fields In STar-forming Region Observations (BISTRO) survey (Project ID: M20AL018), data were acquired in the DAISY mapping mode to ensure fully sampled coverage of the target region.

The data reduction was performed using the pol2map routine within the SMURF package \citep{2005ASPC..343...71B,2013MNRAS.430.2545C} of the Starlink software suite, following the standard BISTRO data reduction pipeline (see \citealt{2021AJ....162..191M}). 
The raw bolometer time-streams were converted into Stokes $I$, $Q$, and $U$ maps using the iterative map-making technique \texttt{makemap}. 
The resulting maps were gridded to a $4''$ pixel scale, sampling the $14.1''$ effective beam \citep{2013MNRAS.430.2534D} at 850 $\mu$m (Nyquist sampling). We applied a standard Flux Conversion Factor (FCF) of $\sim 537$ Jy pW$^{-1}$ beam$^{-1}$ for calibration. 
The total intensity (Stokes $I$) map is derived simultaneously with the polarization data, guaranteeing intrinsic spatial alignment.

We derived the polarization angle ($\theta$) and fraction ($p$) from the Stokes parameters using the relations $\theta = 0.5 \arctan(U/Q)$ and $p = PI/I$. To mitigate the positive bias in polarization measurements due to noise, we adopted the asymptotic estimator for the debiased polarized intensity ($PI$):
\begin{equation}
    PI = \sqrt{Q^2 + U^2 - 0.5(\sigma_Q^2 + \sigma_U^2)}
\end{equation}
where $\sigma_Q$ and $\sigma_U$ represent the uncertainties in Stokes $Q$ and $U$, respectively.
To ensure data quality for the subsequent magnetic field analysis, we selected polarization vectors based on the following criteria: $I/\sigma_I > 10$, $p/\sigma_p > 3$, and $\delta p < 5\%$, where $\sigma_I$ and $\sigma_p$ denote the uncertainties in total intensity and polarization fraction.

\begin{figure}
    \centering
    \includegraphics[width=0.95\linewidth]{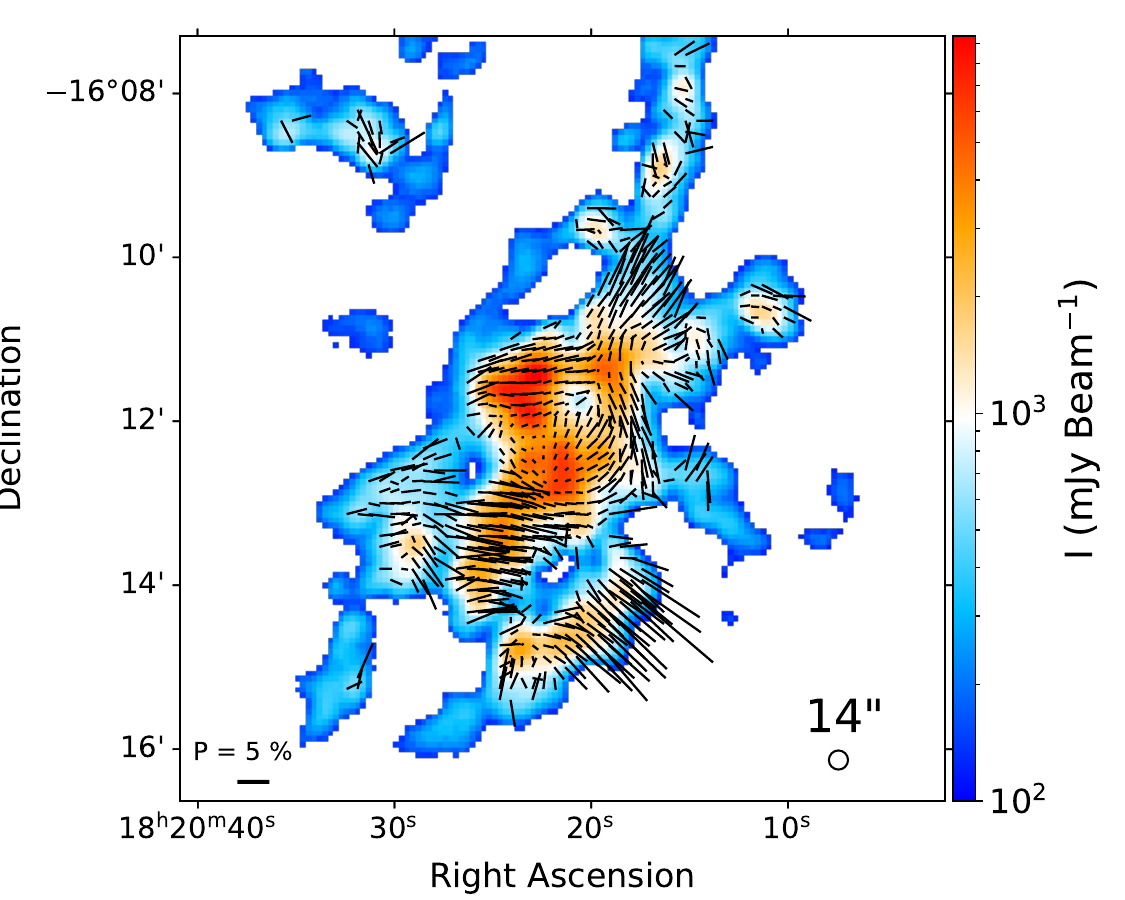}
    \caption{{Magnetic field morphology of M17 SW from 850$\mu$m dust polarization.}
    The background presents the dust continuum emission at 850\,$\mu$m. 
    The black lines show the magnetic field orientations, and their lengths are proportional to the polarization fraction ($p$).}
    \label{fig1}
\end{figure}

\subsection{Spectral Line Data}

To probe the kinematics and physical conditions of the dense gas, we utilize NH$_3$ (1,1) and (2,2) spectral line data from the KEYSTONE (K-band Examinations of Young Stellar Object Natal Environments) survey (Project ID: GBT15B-242; \citealt{2017ApJ...843...63F,2019ApJ...884....4K}). 
Observations were carried out using the Green Bank Telescope (GBT) with the K-band Focal Plane Array (KFPA) and the VEGAS spectrometer.

These observations, targeting the metastable ammonia inversion transitions at 23.69–23.72 GHz, provide a velocity resolution of $\sim 0.11$ km s$^{-1}$ and a beam size of $\approx 30''$.
The NH$_3$ emission, which selectively traces dense gas ($n > 10^3$ cm$^{-3}$) and is less susceptible to depletion than CO, provides robust measurements of the gas systemic velocity ($v_{\rm LSR}$), velocity dispersion ($\sigma_v$), and kinetic temperature ($T_K$). 
Physical parameters were derived by fitting the hyperfine structure of the (1,1) and (2,2) lines simultaneously, using a forward-modeling approach as described in \citet{2019ApJ...884....4K}.



\begin{figure}
    \centering
    \includegraphics[width=0.95\linewidth]{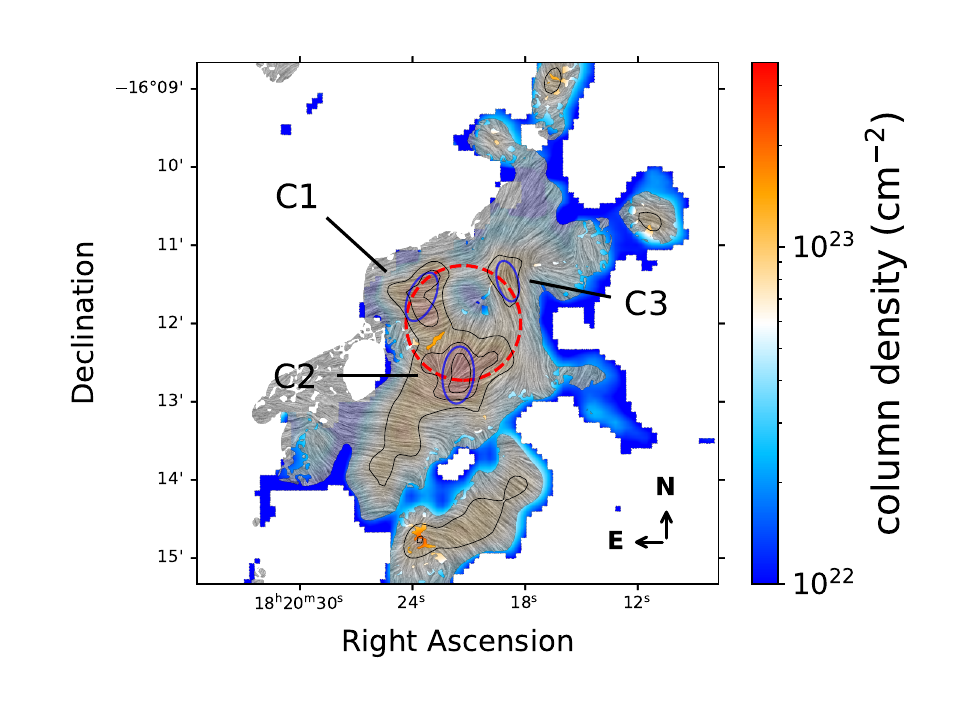}
    \caption{Magnetic field morphology and density structure.
    The LIC map presents the magnetic field morphology from BISTRO dust polarization at scale of 14$''$.
    The background shows the H$_2$ column density at scale of 30$''$ and the black contours display the column density as 1,2,3 $\times$10$^{23}$ cm$^{-2}$, respectively.
    The three blue ellipses show the three dense clumps classified by column density structure and located in a circular shape.}
    \label{fig2}
\end{figure}


\begin{figure}
    \centering
    \includegraphics[height = 6.5cm]{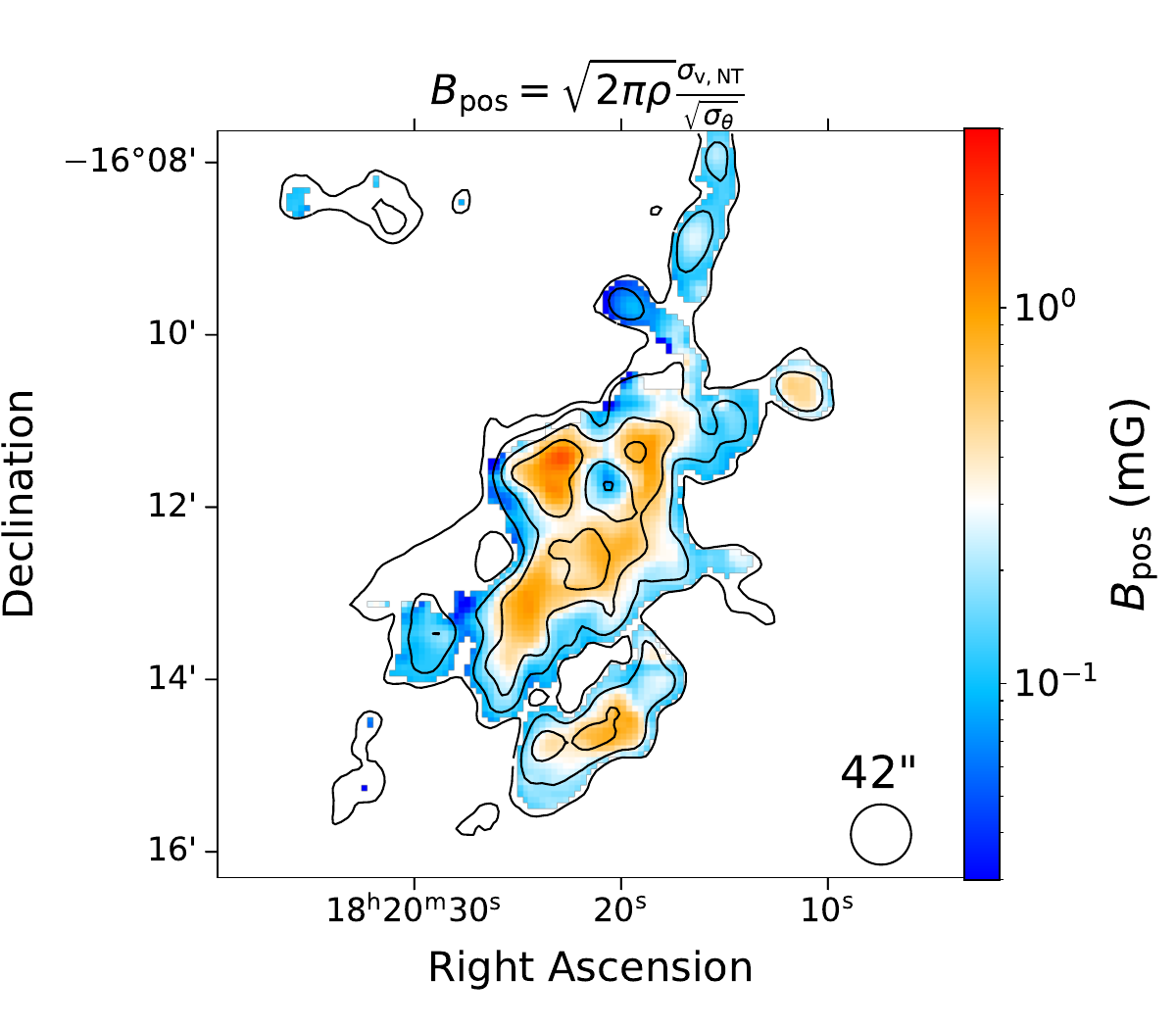}
    \includegraphics[height = 6cm]{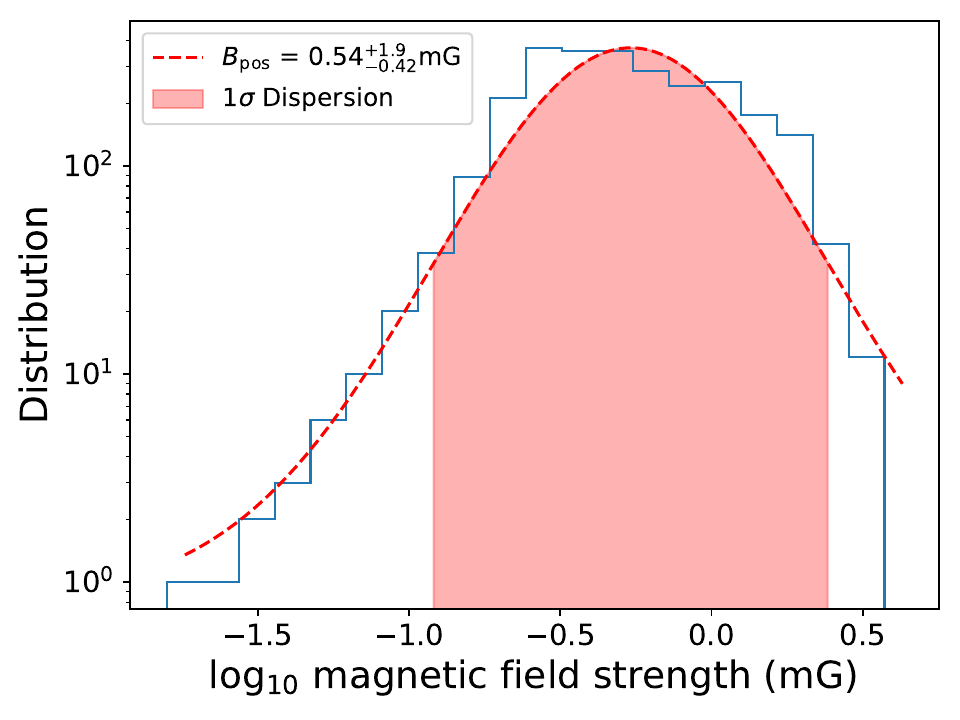}
    \caption{Projected Magnetic field strength distribution derived using the Skalidis-Tassis (ST) method.
    Left panel: Spatial distribution of the plane-of-sky magnetic field strength ($B_{\text{pos}}$) across the M17 SW cloud.
    Black contours show the 850$\mu$m emission from 10$^{2.5}$ to 10$^4$ mJy beam$^{-1}$.  
    right panel: Probability density function (PDF) of the magnetic field strength ($B_{\text{pos}}$) at POS.
    The red dashed line represents a Gaussian fit to the distribution. The shaded pink region corresponds to the $1\sigma$ dispersion range, highlighting the spatial variation of the magnetic field strength. The legend indicates the mean value and the asymmetric dispersion limits derived from the fit.
    All calculations are performed on maps smoothed to a common resolution of 42$''$ to ensure consistency between the polarization and spectroscopic data. }
    \label{fig4}
\end{figure}

\begin{figure}
    \centering
    \includegraphics[width=0.95\linewidth]{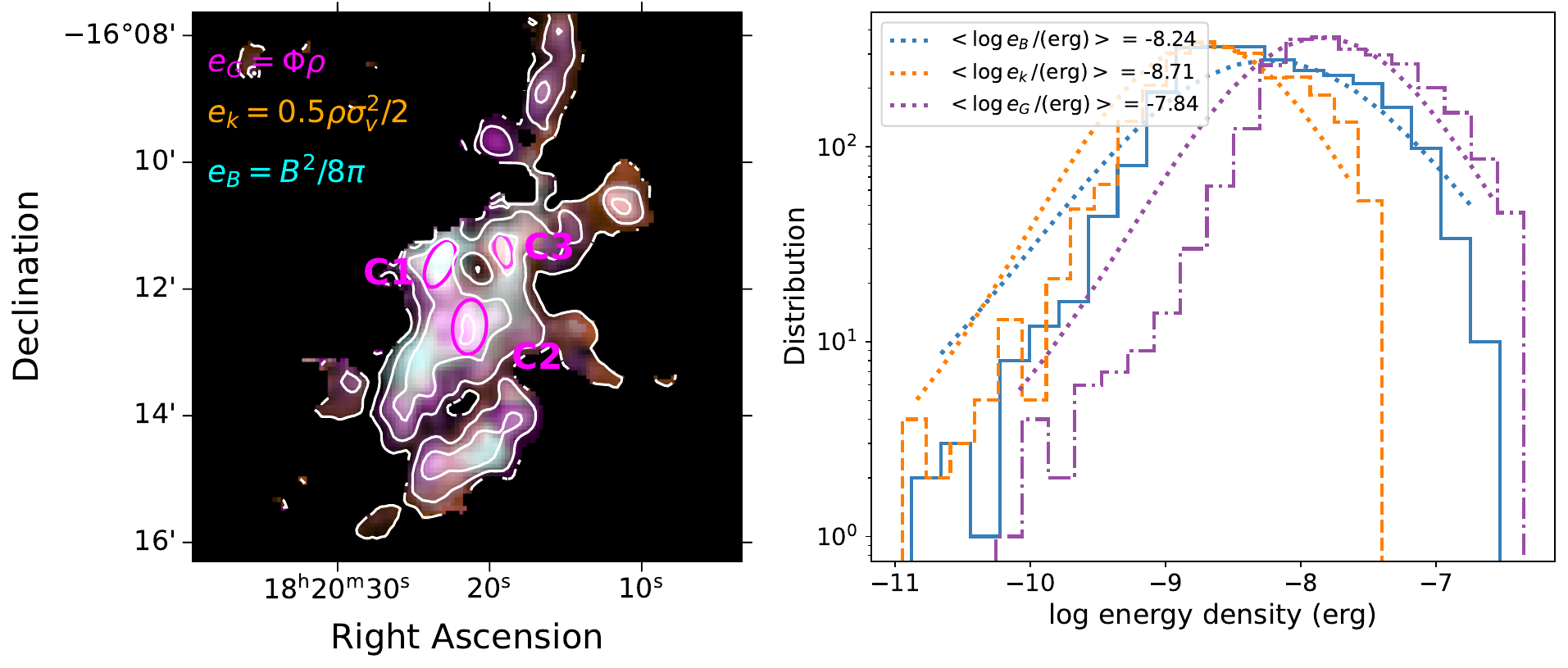}
    \caption{Comparison of gravitational, magnetic, and turbulent energy densities in M17 SW.
    Left Panel: Composite RGB map displaying the spatial distribution of energy densities. 
    Purple represents magnitude of gravitational potential energy density ($|e_G| = \Phi \rho$), orange shows turbulent kinetic energy density ($e_k = 0.5\rho \sigma_v^2$), and cyan indicates magnetic energy density ($e_B = B^2/8\pi$). 
    The magnetic energy density $e_B$ is estimated by applying a statistical correction ($B_{tot} \approx 4/\pi B_{pos}$) to the POS magnetic field strength.
    The background is masked to show only the region with column density measurements ($N_{H_2} > 10^{22}$ cm$^{-2}$). The red ellipses mark the positions of the dense clumps C1, C2, and C3.
    Right Panel: Probability density functions (PDFs) of the three energy components. The dashed lines represent Gaussian fits to the distributions, with the mean values indicated in the legend. 
    This overlap suggests that while gravitational energy is marginally dominant (highest mean), the system is globally in a state of near-equipartition, necessitating a vector analysis to understand the collapse dynamics.
    All energy densities are calculated in cgs units (erg cm$^{-3}$) and are based on maps smoothed to a common resolution of 42$''$. }
    \label{fig5}
\end{figure}

\begin{figure}
    \centering
    \includegraphics[width=0.9\linewidth]{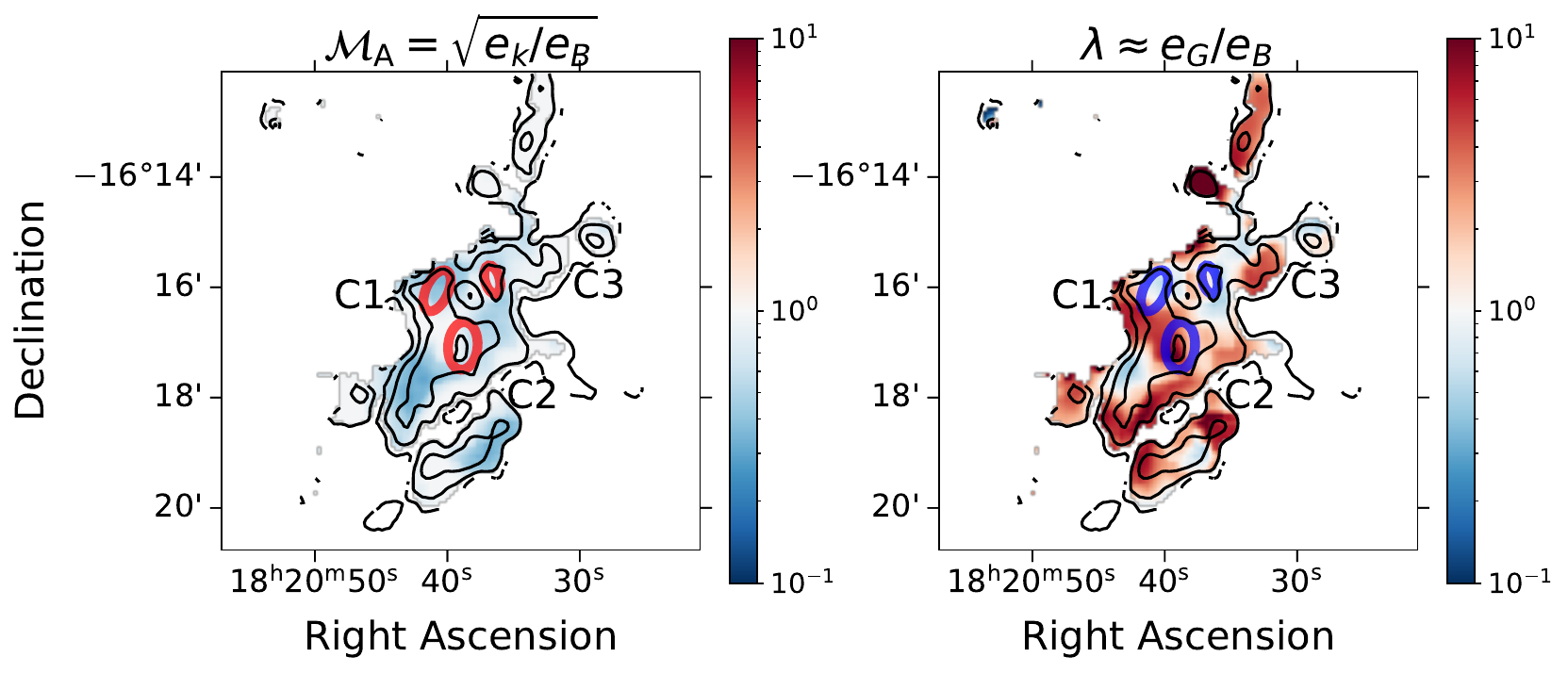}
    \caption{{\bf Distribution of Alfvenic Mach number and mass-to-flux ratio.}
    The backgrounds display the distribution of ${\cal M}_{\rm A}$ and $\lambda$.
    The contours show the structure of column density.}
    \label{figmalambda}
\end{figure}

\section{Result}

\subsection{Magnetic field morphology}

The plane-of-sky magnetic field in M17 SW is globally ordered, yet exhibits significant local deviations coincident with high-density substructures, as shown in Fig.\,\ref{fig1}.
Dominating the morphology is a coherent, arc-like magnetic structure encircling the central region that hosts the three dense clumps C1, C2, and C3 (see Fig.\,\ref{fig2}).
This magnetic arc is open towards the east, facing the ionizing cluster NGC 6618, and connects the clump C3 to the massive clumps C1 and C2.
In the northwestern region, the magnetic field orientation is north-south, parallel to the local filamentary structures. 
In the southeastern region connecting clump C2, the magnetic field orientation reorients to an east-west direction, running perpendicular to the filamentary extended structure. 

\subsection{Derived Physical Parameters}\label{sec3.2}

To quantify the magnetic field strength and energy budget, we derived maps of the H$_2$ column density, volume density ($\rho$), characteristic scale ($r$) and non-thermal velocity dispersion $\sigma_{v, \rm NT}$
The column density map (Fig. 3) was calculated from the $850~\mu$m dust continuum emission and the gas kinetic temperature derived from NH$_3$ hyperfine fitting ( \citep{2019ApJ...884....4K}; see Appendix A for details). 
The volume density and characteristic scale were subsequently determined using the Multi-scale Decomposition Reconstruction (MDR) method (\citealt{2025arXiv250801130Z}; see Appendix B), which spatially decomposes the column density to estimate the local depth of the cloud structure assuming statistical isotropy.
We isolated the non-thermal velocity dispersion from the ammonia linewidths using the relation $\sigma_{v,\rm NT} = \sqrt{\sigma_{obs}^2 - k_B T_{\rm kin}/m_{\rm NH_3}}$, where $\sigma_{obs}$ is the observed velocity dispersion and $m_{\rm NH_3}$ is the molecular mass of ammonia. 
Unlike diffuse tracers (e.g., $^{12}$CO and $^{13}$CO) that require statistical filtering to mitigate extended line-of-sight contamination \citep{2025A&A...700A.256P}, NH$_3$ preferentially traces high-density material. 
This inherently isolates the core dynamics from large-scale flows that could otherwise bias our results \citep{2021MNRAS.506..775P}.
Hereafter, $\sigma_{v,NT}$ is adopted as the tracer of turbulence along the line of sight for all energy density and magnetic field strength calculations.

To quantify the gravitational energy, we calculated the gravitational potential ($\Phi$) directly from the column density map. 
Following the 2D gravitational potential measurement \citep{2023MNRAS.526L..20H}, we solved the Poisson equation in Fourier space assuming a simplified geometry appropriate for the projected data:
\begin{equation}\label{eq2}
\Phi_{k} = - \frac{2\pi G \Sigma_k}{|\mathbf{k}| (1 + |\mathbf{k}| H_{\text{eff}})}
\end{equation}
where $\Sigma_k$ is the Fourier transform of the mass surface density ($\Sigma = \mu m_H N_{H_2}$), $\mathbf{k}$ is the wave vector, and $H_{\text{eff}}$ represents an effective half-thickness (0.3 pc, see Fig.\,\ref{figA2}. 
This approach allows us to recover the 3D potential structure from the projected mass distribution, which is then used to compute the gravitational energy density $e_G = \Phi \rho$ in Sect.\,\ref{sect3.4} and direction of gravitational acceleration $a_G = \nabla\Phi$ (detail in Sect.\,\ref{sect4.2.2}).

\subsection{Magnetic Field Strength}

We estimate the plane-of-sky magnetic field strength ($B_{\text{pos}}$) using the Skalidis-Tassis (ST) method  \citep{2021A&A...647A.186S}, according to the energy equipartition between turbulence and magnetic field of the compressibility of turbulence in molecular clouds. The field strength is calculated as:
\begin{equation}
B_{\text{pos}} = \sqrt{2\pi\rho} \frac{\sigma_{v,NT}}{\sqrt{\sigma_\theta}},
\end{equation}
where $\rho$ is the volume density, $\sigma_{v, \rm NT}$ is the non-thermal velocity dispersion derived from NH$_3$ linewidth, and $\sigma_\theta$ is the angular dispersion of the magnetic field (see Appendix,\ref{Ap.C} for details).

To calculate $\sigma_\theta$, we separated the ordered large-scale magnetic field as background field from the turbulent component by convolving the Stokes $Q$ and $U$ maps with a Gaussian kernel having a full width at half maximum (FWHM) of $42''$, corresponding to approximately $3\times$ the beam size:
\begin{equation}
    \sigma_\theta = \sqrt{\frac{1}{N} \sum (\theta_i - \bar{\theta})^2}
\end{equation}
we excluded pixels with $\sigma_\theta > 25^\circ$ where the dispersion analysis becomes unreliable. 
To ensure physical consistency, all input parameter maps ($\rho$, $\sigma_{v,NT}$) were smoothed to the same $42''$ resolution used for the dispersion analysis (see Fig.\,\ref{fig_Bparam}).

Figure\,\ref{fig4} presents the spatial distribution and statistics of the plane of sky magnetic field strength ($B_{\text{pos}}$). 
The magnetic field strength ranges from 0.1 to 2.4 mG, with a mean value of $\sim 0.54$ mG. 
The highest magnetic field strengths are spatially correlated with the dense clumps C1 and C2. 
To verify the robustness of these estimates, we also performed a calculation using the classical Davis-Chandrasekhar-Fermi (DCF) method. As detailed in Appendix\,\ref{Ap.D}, the DCF method yields a consistent mean magnetic field strength ($\sim 0.69$ mG), confirming the high magnetization of the region regardless of the specific method employed.

\subsection{Energy Density Distribution}\label{sect3.4}

To compare the magnetic field, turbulence, and gravity within a consistent physical framework (same dimension), we assess the dynamical state of M17 SW by comparing the energy densities of the three dominant mechanisms: gravitational potential energy ($e_G = \Phi\rho$), magnetic energy ($e_B = B_{\text{tot}}^2 / 8\pi$), and turbulent kinetic energy ($e_k = \frac{1}{2}\rho \sigma_{v, \text{3D}}^2$).

To ensure dimensional consistency in the energy budget, the total magnetic field strength is estimated by the projected factor ($B_{\text{tot}} \approx \frac{4}{\pi} B_{\text{pos}}$, \citealt{2004ApJ...600..279C}).
As a consistency check, if we decompose this derived $B_{\text{tot}}$ assuming random orientation, the implied line-of-sight component is approximately $\sim 0.4$ mG.
This matches the magnitude of $B_{\text{los}}$ directly measured via HI Zeeman splitting in the M17 envelope (-450/+550 $\mu$G; \citealt{1999ApJ...515..304B}). 
This consistency supports the validity of using the statistically corrected field strength for our dynamical analysis.

Figure\,\ref{fig5} (left panel) displays the composite spatial distribution of these energy densities. 
Gravitational energy ($e_G$) is most concentrated within the dense cores C1–C3, while magnetic energy ($e_B$) appears widespread, showing enhancement along the arc-like structure facing the HII region. 
The probability density functions (PDFs) of the energy densities (Fig.\,\ref{fig5}, right panel) are well-described by log-normal distributions. The derived mean values are $\langle \log e_G \rangle \approx -7.84$, $\langle \log e_B \rangle \approx -8.25$, and $\langle \log e_k \rangle \approx -8.71$ (in units of erg cm$^{-3}$).

While the mean values suggest a hierarchy of $e_G \gtrsim e_B \gtrsim e_k$, the energy density PDFs exhibit overlap, as shown by the $1\sigma$ dispersion widths in Fig.\,\ref{fig5}. 
The differences between the mean energy densities are generally within one order of magnitude (and often within the factor of $\sim 2-3$ uncertainty typical of these estimates). 
This indicates that M17 SW is not dominated solely by gravity but is rather in a state of global near-equipartition, where magnetic fields and turbulence remain dynamically significant competitors to gravitational collapse.

To facilitate comparisons with other star-forming regions  \citep[e.g.,][]{2017ApJ...846..122P,2023ASPC..534..193P}, we recast the energy density analysis into two standard dimensionless parameters: the Alfvénic Mach number ($\mathcal{M}_A$) and the normalized mass-to-flux ratio ($\lambda$). 
The Alfvénic Mach number, representing the ratio of turbulent to magnetic energy, is estimated as:
\begin{equation}
\mathcal{M}_{\rm A} = \frac{\sigma_{v}}{v_A} = \frac{\sigma_{v}}{B_{\text{tot}}/\sqrt{4\pi\rho}} = \sqrt{\frac{(1/2)\rho\sigma_{v}^2}{B_{\text{tot}}^2/8\pi}} \equiv \sqrt{\frac{e_k}{e_B}}.
\end{equation}
Similarly, adopting a flattened geometry, the mass-to-flux ratio approximates the balance between gravity and magnetic fields as 
\begin{equation}
\lambda = \frac{(M/\Phi){\text{obs}}}{(M/\Phi){\text{crit}}} = \frac{\Sigma \cdot A / (B_{\text{tot}} \cdot A)}{1/(2\pi\sqrt{G})} = 2\pi\sqrt{G} \frac{\Sigma}{B_{\text{tot}}} \approx \sqrt{\frac{e_G}{e_B}}.
\end{equation}
Figure\,\ref{figmalambda} presents the spatial distribution of $\mathcal{M}_A$ and $\lambda$ derived from the energy density maps. 
The cloud exhibits a mean $\mathcal{M}_A \approx 0.6$, placing it in a trans- to sub-Alfvénic regime where magnetic tension effectively regulates turbulence. 
However, with a mean $\lambda \approx 2\text{--}3$, the region remains magnetically supercritical, indicating that magnetic support is globally insufficient to halt the gravitational collapse driven by the dominant potential energy.

\begin{figure}
    \centering
    \includegraphics[width=0.95\linewidth]{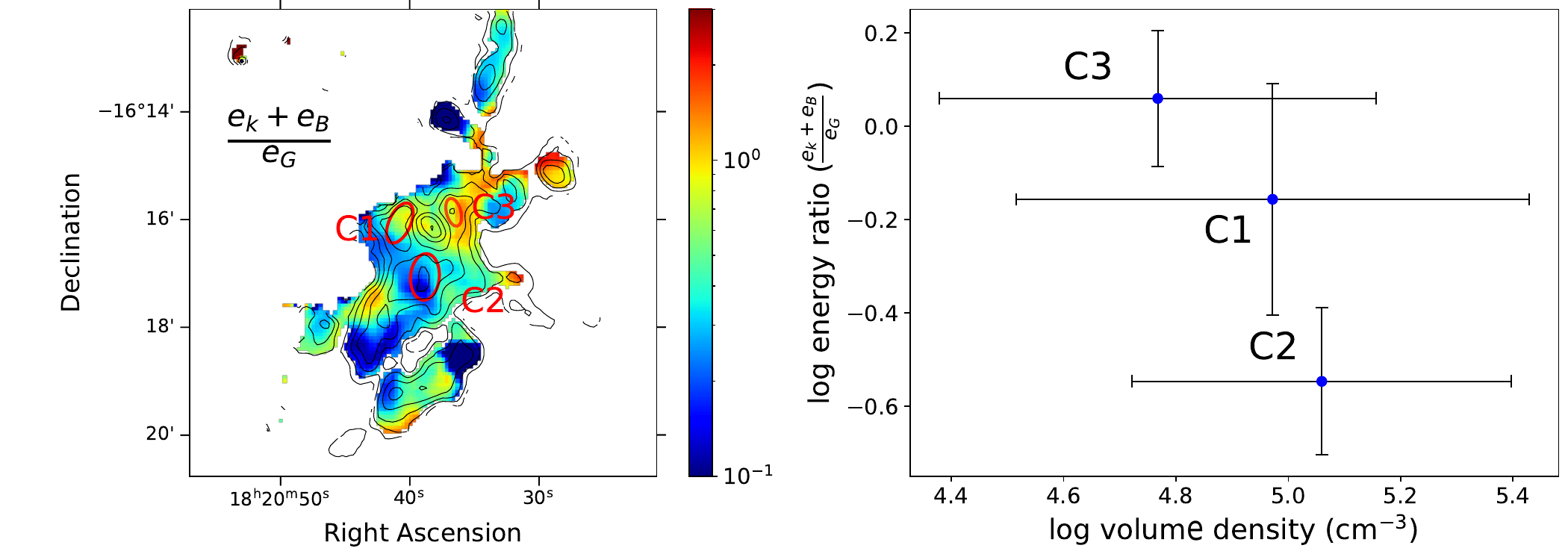}
    \caption{{\bf energy ratio between turbulence, magnetic field and gravity in M17 SW}.
    The left panel shows the energy ratio ($(e_B+e_k)/e_G$ in M17 SW.
    The black contours present the column density structure from 1$\times 10^{22}$ to 3$\times 10^{23.3}$ cm$^{-2}$ with the step of 0.3 dex.
    The right panel shows the distribution of Clump C1, C2, and C3 in the volume density-energy ratio plane. 
    The error bars of clumps present their distribution in the volume density-energy ratio plane. }
    \label{fig8}
\end{figure}

\begin{figure}
    \centering
    \includegraphics[width=0.5\linewidth]{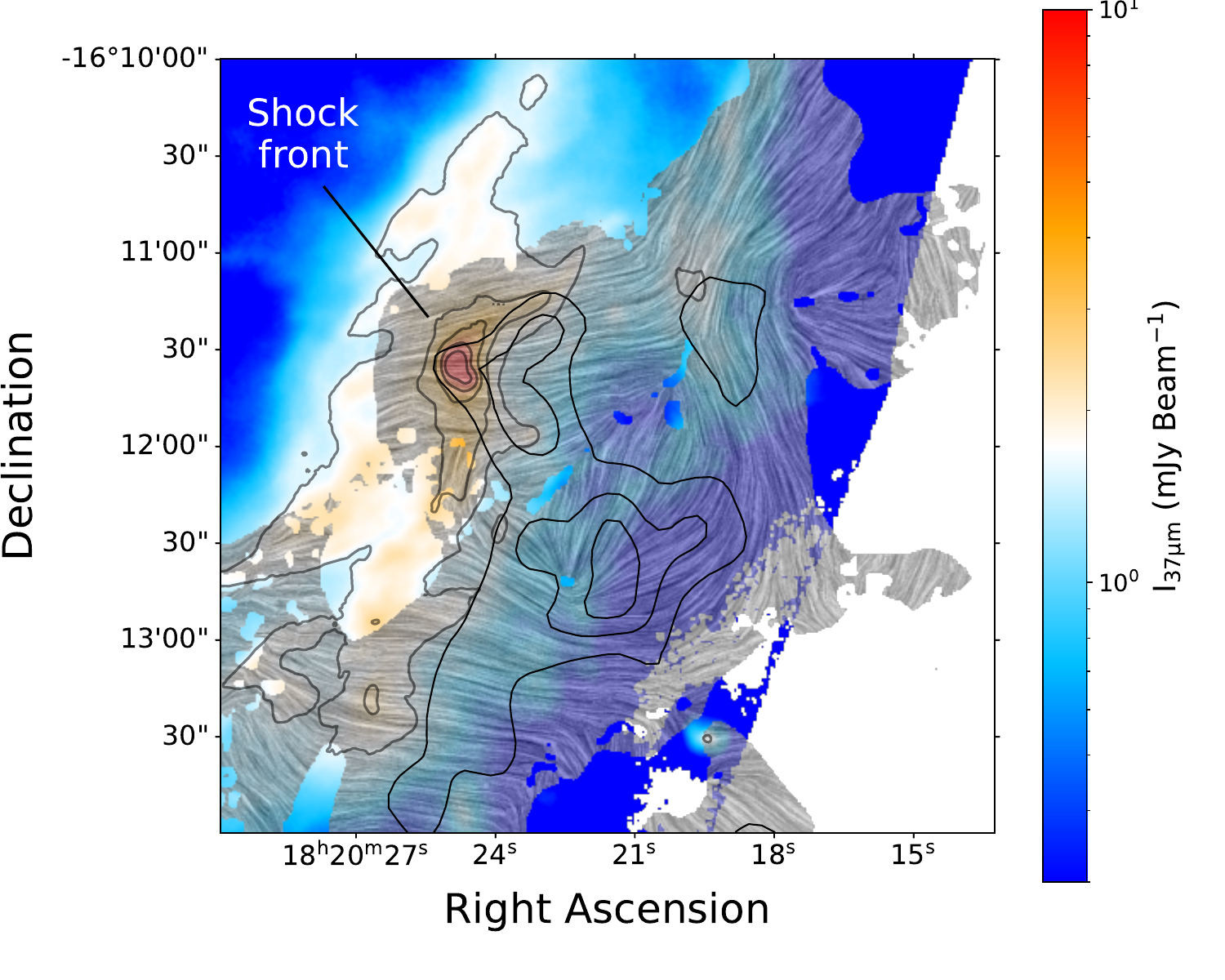}
    \caption{Observational signature of external compression.
    The background image shows 37 $\mu$m emission tracing the warm dust of the PDR and shock front. 
    The overlaid streamlines depict the magnetic field morphology derived from BISTRO data. 
    The strong misalignment of the magnetic field perpendicular to the shock front along the northeastern boundary.
    }
    \label{fig9}
\end{figure}

\begin{figure}
    \centering
    \includegraphics[width=0.95\linewidth]{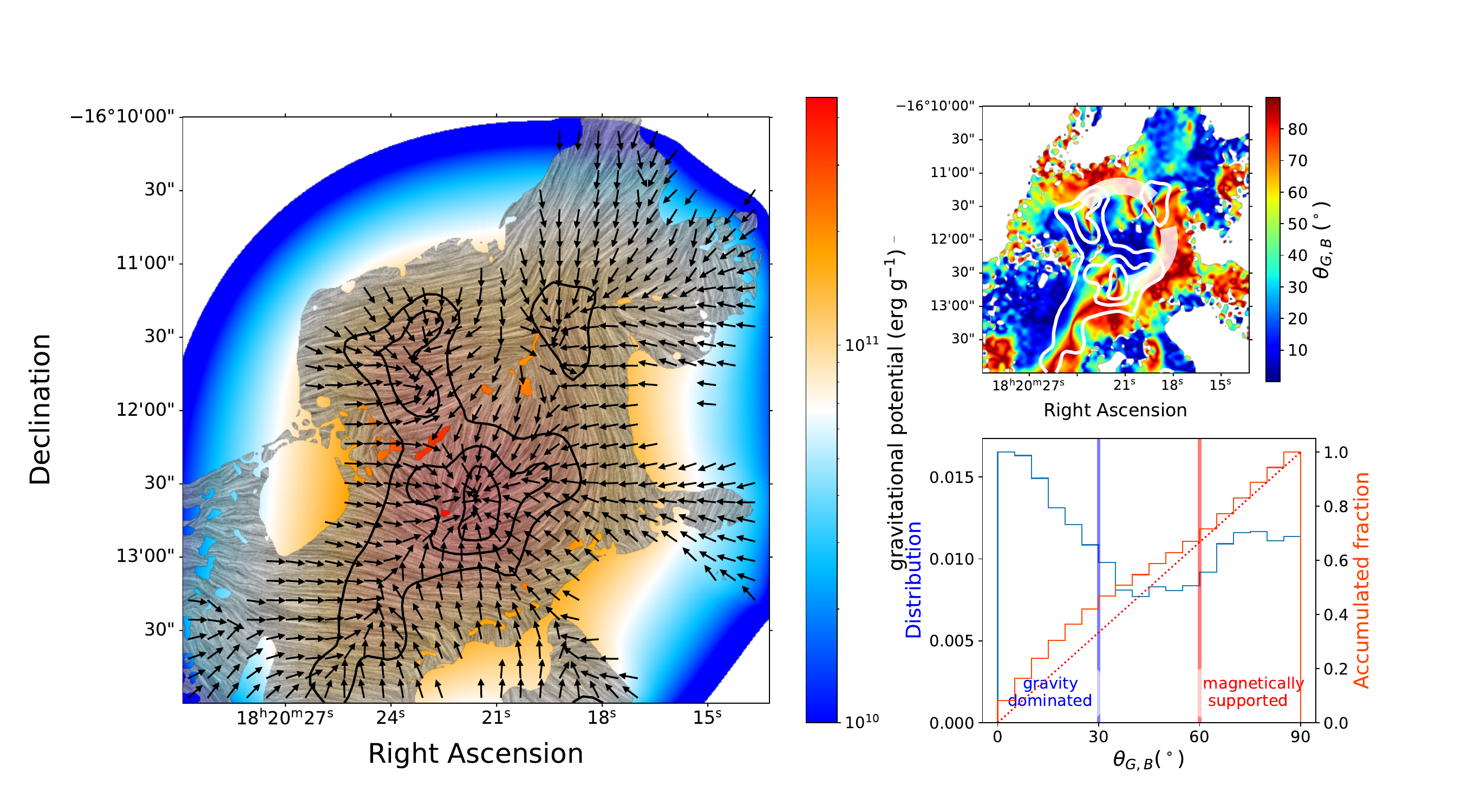}
    \caption{{Alignment between gravity and magnetic field in M17 SW}.
    Left Panel: Gravitational potential overlaid with magnetic field vectors. 
    The LIC map shows the magnetic field orientation, and the black arrows present the direction of local gravity.
    The background is the gravitational potential at POS.
    Top-Right Panel: Spatial distribution of the alignment angle $\theta_{G,B}$. 
    The white dashed paths highlight the "magnetic bridges" connecting C3 to C1/C2.
    The bridge regions clearly exhibit perpendicular alignment ($\theta_{G,B} \approx 90^\circ$, red/yellow), indicating localized magnetic support against radial collapse.
    Bottom-Right Panel: Statistical distribution of alignment angles. 
    The blue line shows the distribution offset angle between magnetic field and gravity.
    The red line presents the accumulation fraction of offset angle, and the dotted line is the random distribution of accumulation fraction.
    The shaded blue region ($\theta_{G,B} < 30^\circ$) represents the gravity-dominated regime which governs the global cloud, while the red region ($\theta_{G,B} > 60^\circ$) corresponds to the magnetically supported regime found within the accretion bridges.}
    \label{fig6}
\end{figure}

\begin{figure}
    \centering
    \includegraphics[width=0.95\linewidth]{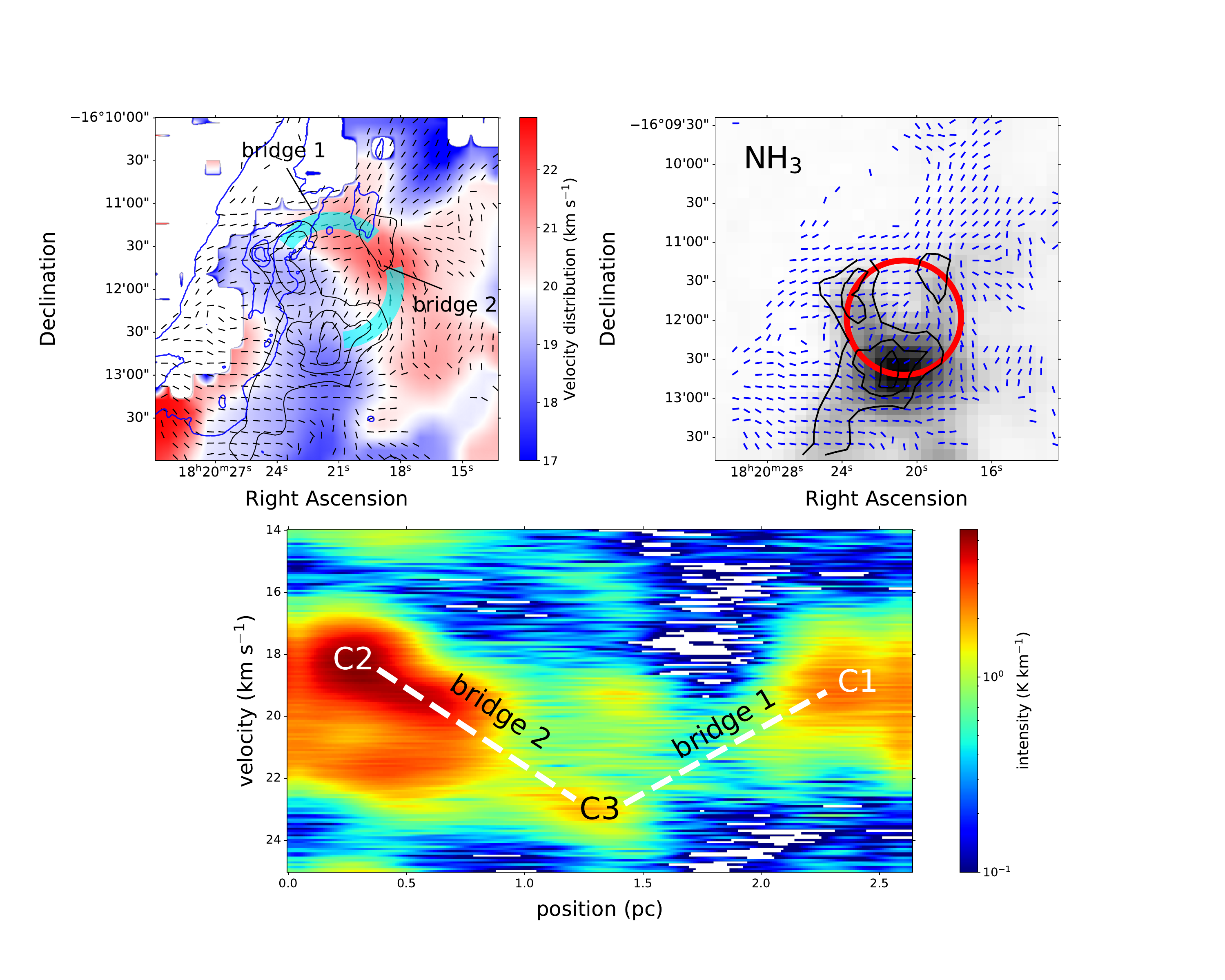}
    \caption{{\bf Kinematic signatures of the accretion bridges.} 
    Top left: Line-of-sight (LOS) velocity field derived from NH$_3$ (color background) overlaid with the magnetic field morphology (black lines) and column density contours (black, starting from $1\times 10^{23}$ to $3\times 10^{23}$ cm$^{-2}$ in steps of $1\times 10^{23}$ cm$^{-2}$). 
    The cyan curves outline the spatial trajectories of the two accretion bridges. 
    Top right: NH$_3$ integrated intensity map. 
    The red circle highlights the central gravitational hub region encompassing the clumps. 
    Bottom: Position-Velocity (PV) diagram extracted along the spine of the accretion bridges (from C1 through C3 to C2, matching the white dashed track). 
    This PV diagram present the velocity gradients characterizing the magnetically-channeled gravitational accretion.
    The white dotted lines show the accretion bridge between C3 and C1, C2 (see Fig.\,\ref{fig2}).}
    \label{fig7}
\end{figure}


\section{Discussion}\label{dis}

\subsection{Scalar Analysis: Global Near-Equipartition and Supercriticality}

The energy density analysis (Sect.\,\ref{sect3.4}) presents a dynamical state defined by competition rather than absolute dominance.
While the mean gravitational potential energy density ($\langle \log e_G \rangle \approx -7.84$) exceeds the magnetic ($\langle \log e_B \rangle \approx -8.25$) and turbulent ($\langle \log e_k \rangle \approx -8.71$) components, the probability density functions exhibit significant overlap (see Fig.\,\ref{fig5}). 
This implies that M17 SW is globally in a state of near-equipartition, where neither magnetic fields nor turbulence can be neglected. 

Figure\,\ref{fig8} maps the spatial variation of this energy balance. 
The ratio $(e_k + e_B)/e_G$ exhibits a clear bimodality: the central dense clumps C1 and C2 are characterized by low ratios ($< 1$), consistent with localized gravitational collapse.  
In contrast, the envelope regions—particularly the arc facing the HII region—show ratios approaching or exceeding unity, indicative of magnetic or turbulent support against gravity.

The dimensionless parameters derived in Sect.\,\ref{sect3.4} reinforce this picture. 
The prevalence of magnetically supercritical yet sub-Alfvénic conditions confirms that while magnetic tension cannot halt the global gravitational collapse, it is sufficiently strong to suppress turbulent disruption. 
Furthermore, since POS measurements underestimate the mass-to-flux ratio \citep{2026A&A...706A..60T}, our derived super-criticality represents a lower limit.
Synthesizing these scalar diagnostics, M17 SW emerges as an unstable system driven by gravity, where the collapse is likely retarded rather than free-fall. 
This suggests a mode of star formation that is globally gravity-dominated but locally regulated by a strong, ordered magnetic field. 
To decipher the mechanics of this regulation, we now turn to a vector analysis of the magnetic field geometry.


\subsection{Vector Analysis: The Squeeze-and-Channel Mechanism}

Vector analysis of magnetic fields, gravity, and gas flows is crucial to reveal the dynamics of physical process, where scalar analysis only studies the energy budget.
The observed morphology supports a "Squeeze-and-Channel" mechanism, resulting from the interaction between magnetic field and external feedback, and gravity.


\subsubsection{External Squeezing: Shock-Driven Alignment}

Figure\,\ref{fig9} shows the magnetic field along the northeastern boundary, adjacent to clumps C1 and C2. 
This region coincides with the PDR shock front traced by 37 $\mu$m warm dust emission  \citep{2020ApJ...888...98L}. 
Along this shock interface and the extended structure southeast of C2, the magnetic field is perpendicular to the shock front.

This perpendicular alignment is consistent with strong external compression (or potentially cloud-cloud collision, \citealt{2023MNRAS.522..503Z}) driven by the expanding HII region.
The shock from the HII region hits the molecular cloud and drags the magnetic field lines along the direction of shock propagation  \citep{2019ApJ...886...17H, 2024ApJ...976..209Z}. 
This indicates that the external HII region shapes the cloud envelope through shock compression  \citep{2003ApJ...590..895W,2015MNRAS.450...10H,2023MNRAS.522..503Z}.



\subsubsection{Accretion Bridges: Magnetically Regulated Gas and support against gravity}\label{sect4.2.2}

The magnetic fields form two curved structures connecting Clump C3 to the massive clumps C1 and C2, respectively (see Fig.\,\ref{fig2}). 
Clump C2 acts as the gravitational center of the region, characterized by the lowest energy ratio and strongest gravitational dominance (see Fig.\,\ref{fig8}).
While these curved structures are faint in the column density map, they are clearly defined in the offset angle map (Fig.\,\ref{fig6}) by their specific alignment: the magnetic field is perpendicular to the local gravity gradient ($\theta_{G,B} \approx 90^\circ$). 
We identify these structures as "accretion bridges", where the magnetic field channels gas flow.

These accretion bridges create a selective support mechanism. 
Across the bridge, magnetic pressure resists gravity, preventing radial collapse and maintaining the channel shape. 
Parallel to the bridge axis, however, magnetic resistance is minimal.
We estimate the enclosed mass of the dense clumps C1 and C2 to be a few $\times 10^3 M_\odot$ based on the column density integration.
A simple dynamic estimate (assuming gravitational acceleration from rest) suggests that this local mass alone might be marginally insufficient to accelerate gas from rest to the observed velocities over the length of the bridge ($\sim 1$ pc).
However, the accretion flow is likely driven by the global gravitational potential of the entire M17 SW cloud (total mass $> 10^4 M_\odot$, derived from column density; see Fig.\,\ref{A1}), rather than by the local clumps alone. As gravity is a long-range force, the massive envelope contributes significantly to the acceleration.
Furthermore, the gas likely enters these magnetic channels with non-zero initial momentum derived from the external shock compression (HII region expansion).
The moderate local mass (Mass of Clumps C1 and C2 $\sim 3 \times 10^3 M_\odot$) is actually critical for this regulation; if the local mass were higher ($> 10^4 M_\odot$), the local gravitational energy would likely overwhelm the magnetic support, disrupting the bridge geometry and leading to free-fall collapse.

This mechanism allows gas to move toward the gravity center. 
We extracted a Position-Velocity (PV) diagram along the spatial spine of the accretion bridges connecting C1, C3, and C2. 
As shown in the bottom panel of Figure\ref{fig7}, the NH$_3$ emission exhibits a continuous and coherent velocity gradient of $\sim$ 4-6 km s$^{-1}$ pc$^{-1}$ along these specific tracks.
Driven by the strong gravity of C2 (and the global potential, see Fig.\,\ref{fig6}), gas flows unimpeded from C3 through magnetically supported bridges (see Fig.\,\ref{fig8}). 
The magnetic field acts as a structural guide, channeling this shock-injected, gravity-assisted flow directly onto the potential well of C2, preventing the gas from dispersing or stalling.

\subsection{Synthesis: A Unified Physical Picture}

Combining the energy budget (scalar) and alignment (vector) analyses, we propose a physical scenario for M17 SW, characterized as a Squeeze-and-Channel process:
\begin{itemize}
    \item External Compression: The interaction begins at the northeastern boundary. The expanding HII region (NGC,6618) drives a strong shock into the molecular cloud. 
    This event, likely a cloud-cloud collision, compresses the gas and drags the magnetic field lines, aligning them perpendicular to the shock front. This external force defines the high-density envelope of the cloud.
    \item Magnetic Channeling in Accretion Flows: Distinct from the shock front, the magnetic field organizes into curved accretion bridges connecting the gas reservoir (C3) to the gravitational center (C2). 
    These accretion bridges provide selective support: magnetic pressure prevents the gas from collapsing radially, maintaining the bridge structure, while allowing free movement along the bridge axis.
    \item Gravitational Drive: Although the system is near energy equipartition, gravity remains the primary driver. It pulls gas from C3 through these magnetically supported bridges, accumulating mass onto the massive clump C2.
\end{itemize}

In summary, star formation in M17 SW is globally driven by gravity but locally regulated by magnetic fields. 
The external shock shapes the initial condition, the magnetic field defines the accretion pathways, and gravity powers the final mass accumulation.

\section{Summary}

In this work, we study the roles of magnetic fields, gravity, and turbulence in the massive star-forming region M17 SW by energy density analysis and kinetic analysis, combining high-resolution 850 $\mu$m dust polarization observations (BISTRO/JCMT) with NH$_3$ spectral line data.
Our main findings are:

1. Magnetic Field Morphology $\&$ Strength: \\
The magnetic field exhibits a striking ring-like structure surrounding the three massive clumps (C1, C2, C3). The POS magnetic field strength, measured using the Skalidis-Tassis method, ranges from 0.1 to 2.9 mG, with a mean strength of 0.54 mG.

2. Energy Budget $\&$ Near-equipartition System: \\
Analysis of the energy densities reveals a clear hierarchy: gravity dominates ($\langle e_G \rangle \approx 10^{-7.84}$ erg cm$^{-3}$), followed by magnetic energy ($\langle e_B \rangle \approx 10^{-8.25}$ erg cm$^{-3}$) and then turbulent energy ($\langle e_k \rangle \approx 10^{-8.71}$ erg cm$^{-3}$). 
The cloud is magnetically supercritical ($\lambda \gtrsim 1$) and sub-Alfvénic state (${\cal M}_{\rm A} \lesssim 1$), where the gravity dominates the main physical process and the magnetic field also plays an important role.
Due to energy densities of gravity, magnetic field, and turbulence on one order of magnitude, the M17\,SW could be close to a near-energy-equipartition system.

3. Accretion Bridges $\&$ Channeling Flow: \\
We identified curved "accretion bridges" connecting Clump C3 to the massive clump C2. 
In these bridges, the magnetic field is perpendicular to gravity. 
This configuration provides support against radial collapse but allows gas to flow freely along the bridge axis. 
Velocity gradients confirm that gas is flowing from C3 onto C2. 
This flow is likely powered by the global gravitational potential and guided by the magnetic field lines.

4. External Shock: \\
The northeastern boundary shows signs of strong external compression from the adjacent HII region. 
Here, the magnetic field is perpendicular to the shock front, consistent with magnetic field lines being dragged by a shock. This external force defines the dense envelope of the cloud.

Star formation in M17 SW is globally driven by gravity but locally regulated by the magnetic field. 
The external shock shapes the cloud's initial structure, the magnetic field defines the accretion channels, and gravity powers the final accumulation of mass onto the cores.

\begin{acknowledgments}
M.Z. acknowledges support from the National Natural Science Foundation of China (NSFC) grant No. 12503029.
K.Q. is supported by NSFC grant No. 12425304, the National Key R\&D Program of China (Nos. 2023YFA1608204 and 2022YFA1603103), and the grant from the China Manned Space Project.
E.C. acknowledges the support from Core Research Grant (CRG; sanction order number CRG/2023/008710) awarded by Anusandhan National Research Foundation (ANRF) under Science and Engineering Research Board (SERB), Govt. of India.
\end{acknowledgments}

\bibliographystyle{aasjournal}
\bibliography{reference}

\appendix

\section{Column density}\label{Ap.A}

\begin{figure}
    \centering
    \includegraphics[width=0.95\linewidth]{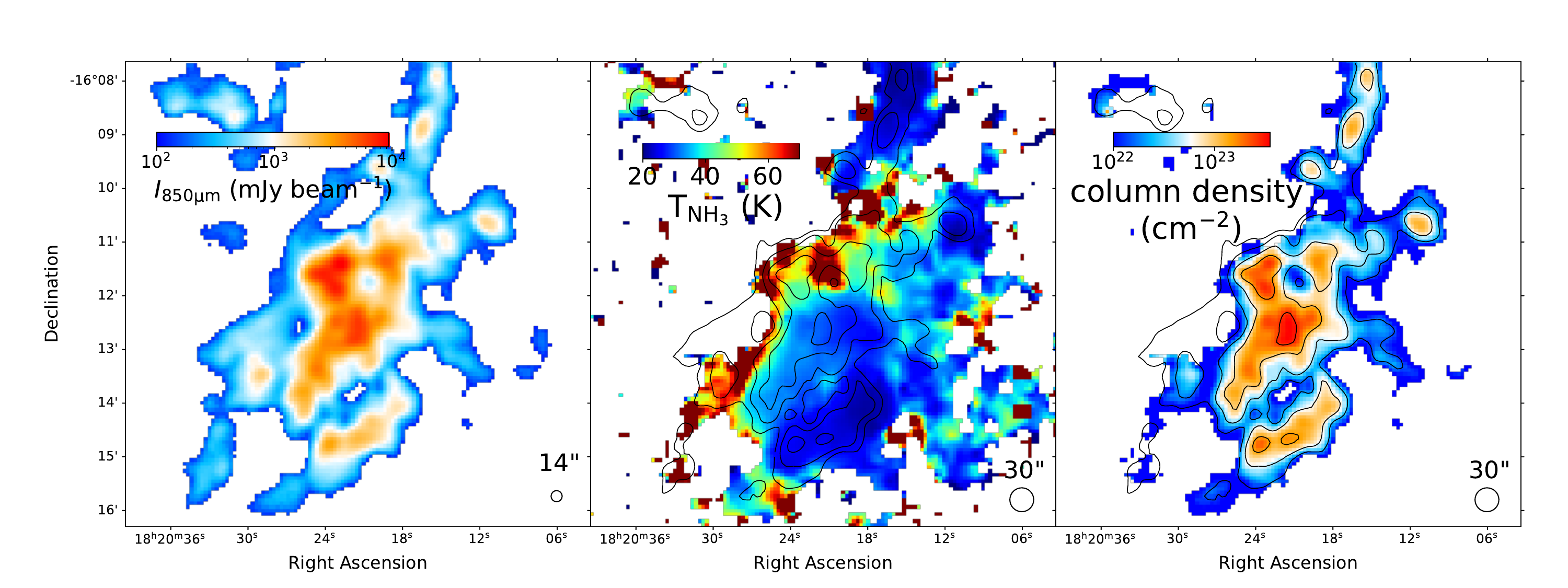}
    \caption{{\bf Distribution of column density is measured by  850$\mu$m emission and dense gas temperature (NH$_3$).}
    The panels present the distribution of 850$\mu$m emission, gas temperature from NH$_3$  \citep{2017ApJ...843...63F} and H$_2$ column density, from left to right.}
    \label{A1}
\end{figure}

The column density is a fundamental physical parameter to describe the mass distribution of interstellar medium on the plane of sky.
In general, it can be calculated by the blackbody model and the intensity of the continuum.
In this work, we use the intensity map of 850$\mu$m emission, observed by JCMT Scuba-2, to measure the column density map.
According to the blackbody function, the column density traced by warm and cold dust can be calculated as:
\begin{equation}\label{eqSigma}
    \Sigma = \frac{R S_\nu}{B_\nu(T_{\rm dust}) \kappa_\nu}
\end{equation}
where \( R \) is the gas-to-dust mass ratio (\(\approx 100\)), \( S_\nu \) is monochromatic flux at frequency $\nu$, \( B_\nu(T_{\rm dust}) \) is the Planck function at the dust temperature \( T_{\rm dust} \), and \( \kappa_\nu \) is the dust opacity (\(\approx  1.48 \, \mathrm{cm^2 \, g^{-1}}\) at 850\,$\mu$m, based on the MRN model with thin ice mantles after \( 10^5 \) years; \citealt{1994A&A...291..943O}).
The dust temperature is similar to the temperature of dense molecular gas, probed by NH$_3$ from GBT  \citep{2017ApJ...843...63F}, which assumes the similar temperature of gas and dust.
To uniform the spatial resolution, the 850$\mu$m emission is smoothed as $30''$, the same resolution of NH$_3$.
Then, using the Eq.\,\ref{eqSigma}, the column density is calculated at the scale of $30''$ (see Fig.\,\ref{A1}).

\section{mapping Volume density and  characteristic scale}\label{Ap.B}

In this work, we derive the volume density ($\rho$) and characteristic scale ($l_c$) distributions using the Multi-scale Decomposition Reconstruction (MDR) method \citep{2026ApJ...997..345Z}, which is built upon the constrained diffusion algorithm \citep{2022ApJS..259...59L}.

Traditional density estimations assume a uniform line-of-sight (LOS) thickness across the entire cloud. 
Such a rigid assumption overestimates volume densities in extended diffuse envelopes and underestimates them in compact dense cores. 
The MDR method overcomes this structural bias by dynamically extracting the local structural width from the column density map \citep{2026ApJ...997..345Z}.

\begin{figure}
    \centering
    \includegraphics[width=0.95\linewidth]{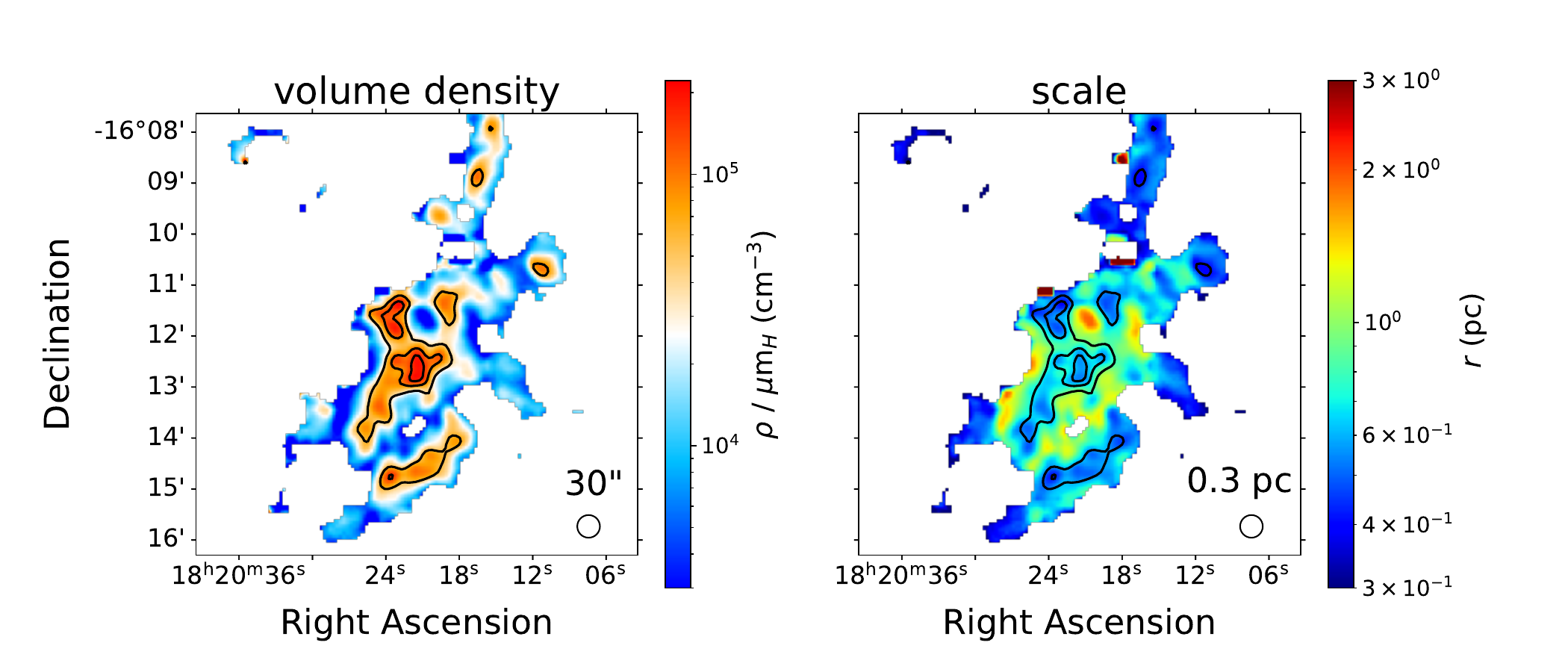}
    \caption{{\bf Decompose the column density as volume density and characteristic scale map.}}
    \label{figA2}
\end{figure}

The MDR method operates by: (1) decomposing the column density map into scale-dependent components; (2) computing the intensity-weighted characteristic scale ($l_c$) at each position; and (3) deriving the volume density under the assumption of local statistical isotropy. Here, the LOS effective thickness is locally approximated by the POS structural width ($l_{\rm t, LOS} \approx l_c$), allowing the thickness to adapt to the hierarchical structure of the cloud \citep{2019MNRAS.490.3061V}. 
Fig.\,\ref{figA2} presents the resulting POS distributions of the volume density and characteristic scale in M17 SW.

While macroscopic external compression (e.g., from the adjacent HII region at the northeastern boundary) might introduce local geometric flattening, 
This variable-thickness framework is fundamentally superior to a constant-thickness assumption. 
\citealt{2026ApJ...997..345Z} using the 3D MHD validations demonstrate that this statistically isotropic scaling inherently captures the core-envelope density contrast and robustly constrains the global density uncertainties within a factor of $\sim 2$, fulfilling the requirements for dynamical analyses in this work.

\section{physical parameters for estimating magnetic field strength}\label{Ap.C}

\begin{figure}
    \centering
    \includegraphics[width=0.95\linewidth]{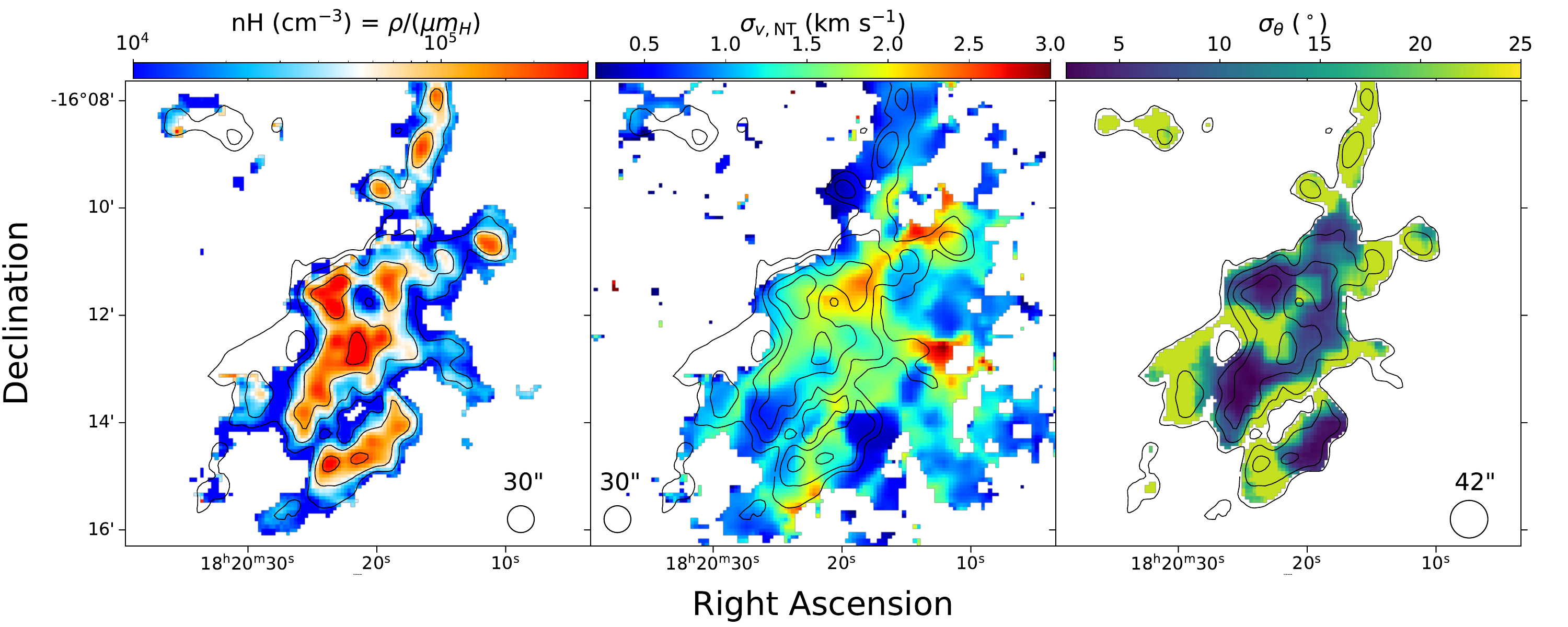}
    \caption{{\bf Distribution of volume density, non-thermal velocity dispersion, and magnetic field orientation angle dispersion.}}
    \label{fig_Bparam}
\end{figure}

In this work, the physical parameter for measuring the magnetic field strength only utilizes two datasets to minimize systematic errors from observation, the observation of 850$\mu$m dust polarization and NH$_3$ (1-1) (2-2).
H$_2$ volume density is obtained by decomposing column density using the Multiscale Decomposition Reconstruction (detail show in Ap.\,\ref{Ap.B},\citealt{2025arXiv250801130Z}), which the column density come from 850$\mu$m continuous and gas temperature from NH$_3$.
The volume density, non-thermal velocity dispersion, and magnetic field orientation angle dispersion are the necessary parameters to estimate magnetic field by turbulence- magnetic energy equipartition  \citep{1951PhRv...81..890D,1953ApJ...118..113C,2004ApJ...600..279C,2021A&A...647A.186S,2021A&A...656A.118S,2021ApJ...919...79L}.
The detail of distribution is shown in Fig.\,\ref{fig_Bparam}

\section{COMPARISON OF ST AND DCF METHODS}\label{Ap.D}

\begin{figure}
    \centering
    \includegraphics[width=0.9\linewidth]{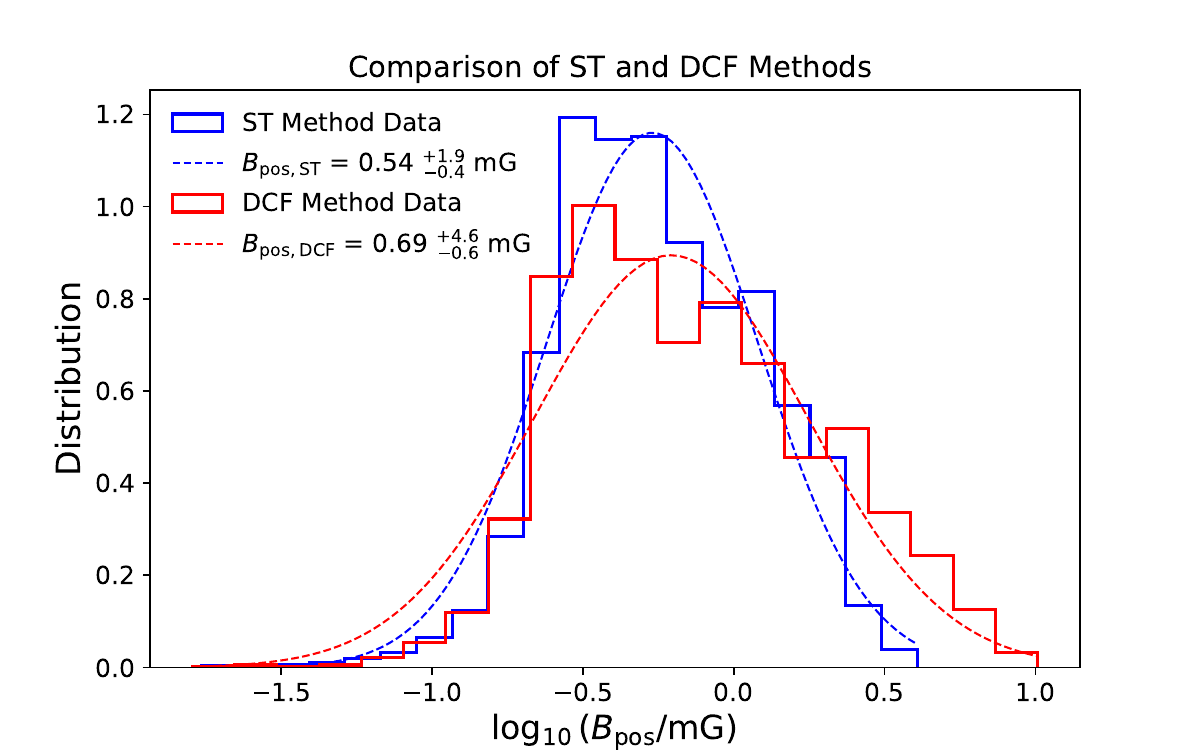}
    \caption{Comparison of POS magnetic field strength estimates using the ST and DCF methods.
    Histograms of the POS magnetic field strength ($B_{\text{pos}}$) derived from the ST method (blue) and the DCF method (red). The red dashed lines show the Gaussian fits. The ST method yields a mean of $\sim 0.54$ mG, while DCF yields $\sim 0.69$ mG.}
    \label{figA3}
\end{figure}

While the main text utilizes the Skalidis-Tassis (ST, \citealt{2021A&A...647A.186S,2021A&A...656A.118S}) method due to its consideration of compressible turbulence modes, the Davis-Chandrasekhar-Fermi (DCF, \citealt{1951PhRv...81..890D,1953ApJ...118..113C,2022ApJ...925...30L}) method remains a standard benchmark in the field. 
To ensure our conclusions regarding the energy budget are model-independent, we compare the results of both techniques. 
The DCF field strength is calculated as:
\begin{equation}
    B_{\text{pos, DCF}} = Q\sqrt{4\pi\rho} \frac{\sigma_v}{\sigma_\theta} 
\end{equation}
where we adopt a correction factor of $Q=0.4$ \citep{2021ApJ...919...79L}.
Figure\,\ref{figA3} presents the comparison. The ST method yields a mean POS magnetic field strength of $B_{\text{pos,ST}} = 0.54^{+1.9}_{-0.4}\text{mG}$, while the DCF method yields $B_{\text{pos, DCF}} = 0.7^{+4.6}_{-0.6} \text{mG}$. 
The systematic difference is approximately 25$\%$, which is within the typical uncertainty range for polarization-based estimates. 

\section{Uncertainties in Gravitational Potential from Geometric Assumptions}

\begin{figure}
    \centering
    \includegraphics[width=0.5\linewidth]{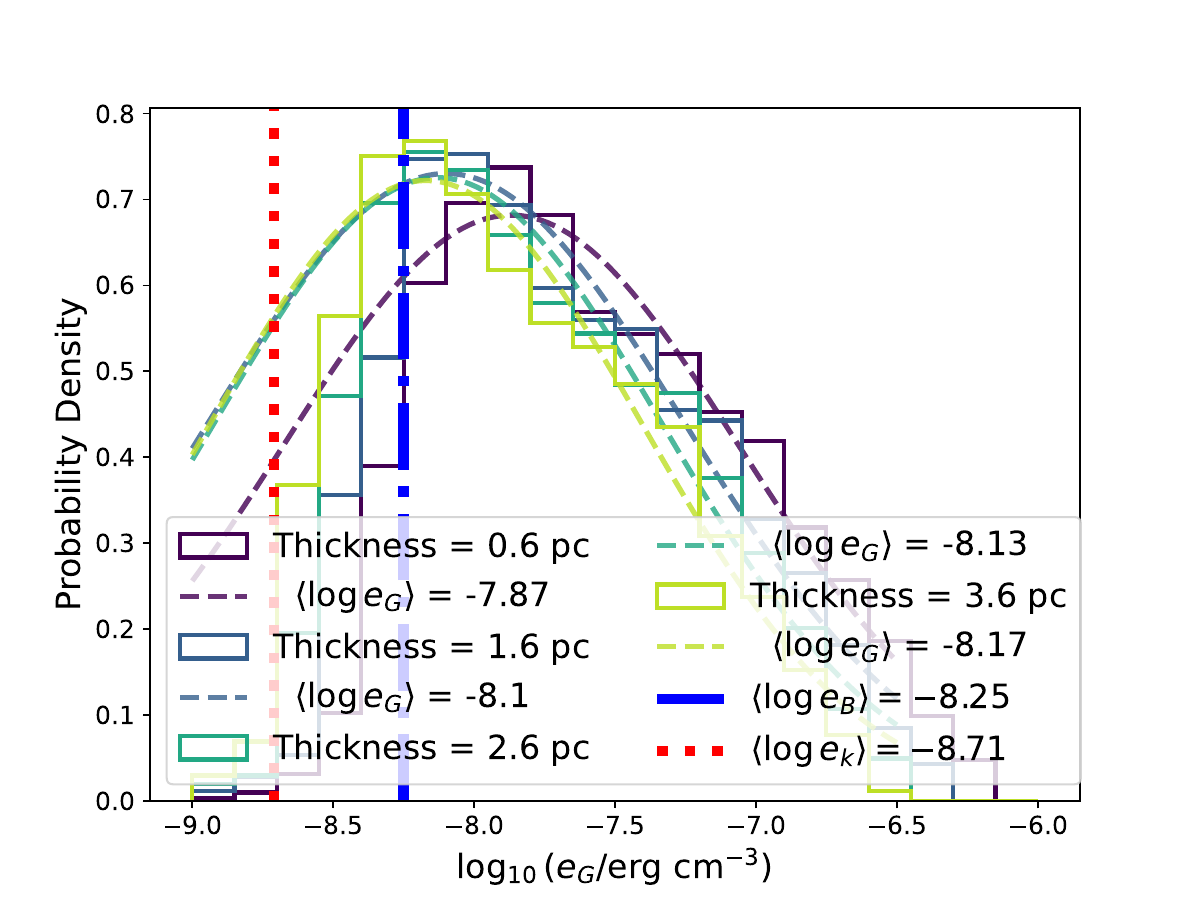}
    \caption{Probability density distributions of the gravitational energy density ($e_G$) under varying assumptions for the line-of-sight cloud thickness. The simulated effective thicknesses range from 0.6 pc to 3.6 pc. Dashed curves represent Gaussian fits to each distribution, with their respective mean values ($\langle \log e_G \rangle$) annotated in the legend. For direct comparison, the mean magnetic energy density ($\langle \log e_B \rangle = -8.25$) and turbulent kinetic energy density ($\langle \log e_k \rangle = -8.71$) are denoted by the blue dash-dotted and red dotted vertical lines. The gravitational energy density unequivocally remains the dominant component across all tested geometric configurations.}
    \label{figE1}
\end{figure}

We estimated the gravitational potential assuming a flattened slab geometry with a fixed effective half-thickness (Equation\,\ref{eq2}) in section\,\ref{sec3.2}. 
To quantify the uncertainties introduced by this simplified geometric assumption, we tested the sensitivity of the derived gravitational energy density ($e_G$) to the assumed line-of-sight extent of the cloud. 
We varied the total effective cloud thickness across a broad parameter space, ranging from 0.6 pc to 3.6 pc.

Figure\,\ref{figE1} presents the resulting probability density functions of $e_G$ for these varying geometric scenarios. 
A larger assumed thickness naturally distributes the observed column density over a longer sightline, thereby diluting the local volume density.
This dilution slightly reduces the absolute magnitude of the gravitational potential and its corresponding energy density. 
Consequently, the mean value of the gravitational energy density ($\langle \log e_G \rangle$) shifts gradually from $-7.69$ to $-8.07$ as the cloud thickness increases from 0.6 pc to 3.6 pc.

Despite this anticipated geometric variance, the gravitational energy density consistently governs the energy budget. 
Across all tested spatial configurations, $\langle \log e_G \rangle$ remains strictly higher than both the mean magnetic energy density ($\langle \log e_B \rangle = -8.25$) and the turbulent kinetic energy density ($\langle \log e_k \rangle = -8.71$). 
This exploration confirms the primary dynamical conclusion. 
The characterization of M17 SW as a globally gravity-dominated system is remarkably robust, remaining entirely unaffected by uncertainties in the unconstrained line-of-sight geometry.

\begin{figure}
    \centering
    \includegraphics[width=0.5\linewidth]{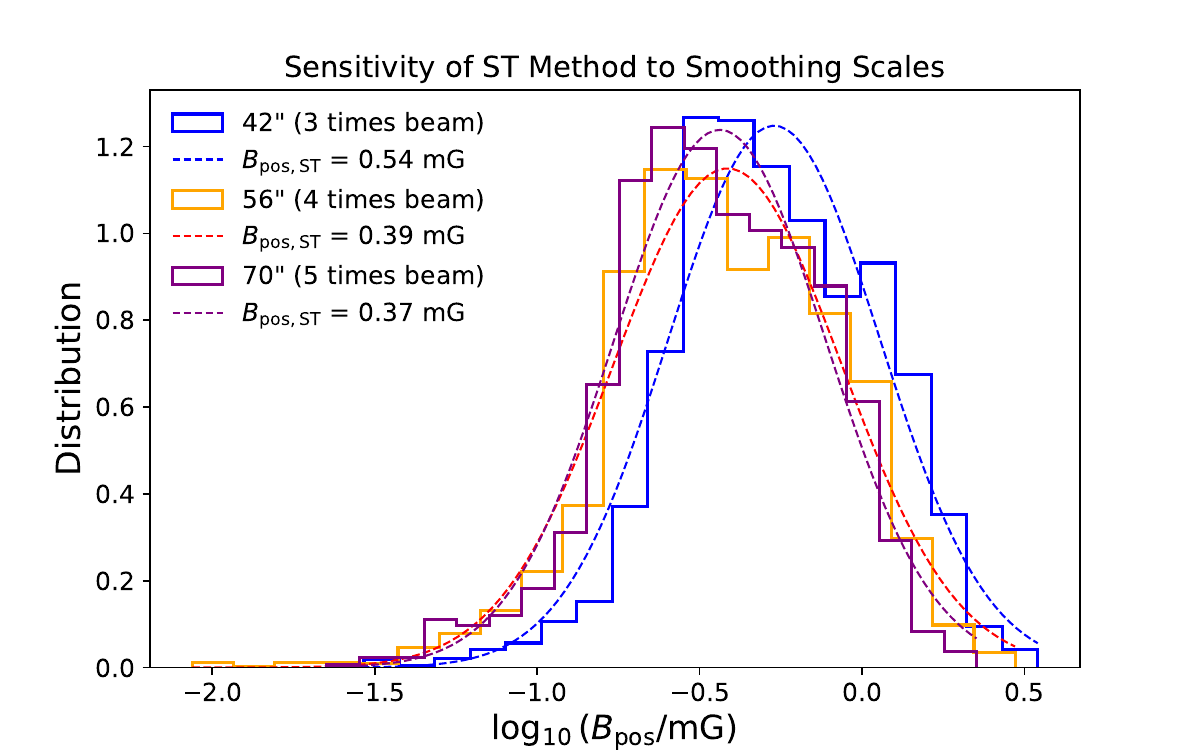}
    \caption{Probability density distributions of the plane-of-sky magnetic field strength ($B_{\rm pos}$) derived using the ST method with varying spatial smoothing kernels: 42'' (blue), 56'' (orange), and 70'' (purple).
 The dashed curves represent Gaussian fits, with the mean field strengths annotated. The inferred field strength slightly decreases and stabilizes at larger smoothing scales, which further reinforces the gravity-dominated dynamical state of the cloud.}
    \label{figF1}
\end{figure}

Extracting the turbulent magnetic field in the ST and DCF methods requires spatial smoothing to filter out the large-scale ordered field \citep{1951PhRv...81..890D,1953ApJ...118..113C,2021ApJ...919...79L,2022ApJ...925...30L}. 
The resolution of the background order magnetic field should distinctly differ to observed beam size.
We adopt a physical lower limit of 42$''$, corresponding to three times (half dex) 14$''$ beam size.

To assess the sensitivity of our results to this parameter, we recalculated the magnetic field strength using smoothing kernels of 42'', 56'', and 70'' (representing 3$\times$, 4$\times$, and 5$\times$ the beam size). 
Figure\,\ref{figF1} presents the resulting distributions. Increasing the kernel size from 42'' to 56'' slightly reduces the mean $B_{\rm pos}$ from 0.54 mG to 0.39 mG, after which the value stabilizes.

This slight decrease in $B_{\rm pos}$ at larger smoothing scales perfectly aligns with fundamental physical expectations. 
A larger spatial kernel effectively averages the parameters over a broader physical extent, inherently incorporating a higher fraction of the lower-density envelope gas. According to the well-established magnetic field-density ($B-\rho$) relation, magnetic field strength intrinsically scales with gas density \citep{2010ApJ...725..466C,2015Natur.520..518L,2024ApJ...967...18Z}. 
Therefore, measuring a weaker mean $B_{\rm pos}$ at a larger, less-dense macroscopic scale is a direct physical manifestation of the cloud's hierarchical structure, rather than a methodological artifact.

\end{document}